\pdfoutput=1

\documentclass[conference]{IEEEtran}
\IEEEoverridecommandlockouts
\usepackage[misc]{ifsym}
\usepackage[table]{xcolor}  
\usepackage{tikz}
\usepackage{amsmath}
\usepackage{graphicx}
\usepackage{booktabs}
\usepackage{amssymb}
\usepackage{latexsym}	
\usepackage{array}		
\usepackage{multirow}
\usepackage{subfigure}
\usepackage{algorithm}
\usepackage{algpseudocode}
\usepackage[font={small,bf}]{caption}  
\usepackage{paralist}   
\usepackage{xspace}
\usepackage{color}
\usepackage{adjustbox}
\usepackage{balance}
\usepackage{wasysym}
\usepackage{rotating}   
\usepackage{listings}
\usepackage{flushend}  
\usepackage{tabu}
\usepackage{amssymb}
\usepackage{pifont}
\usepackage{framed}

\usepackage{url}            
\usepackage{hyperref}

\usepackage{xpatch}
\makeatletter
\chardef\TPT@@@asteriskcatcode=\catcode`*
\catcode`*=11
\xpatchcmd{\threeparttable}
  {\TPT@hookin{tabular}}
  {\TPT@hookin{tabular}\TPT@hookin{tabu}}
  {}{}
\catcode`*=\TPT@@@asteriskcatcode
\makeatother

\usepackage{cite}

\ifCLASSINFOpdf
\else
\fi
\newcommand*\circled[1]{\tikz[baseline=(char.base)]{
            \node[shape=circle,fill,inner sep=1pt] (char) {\textcolor{white}{#1}};}}

\newcommand{\red}[1]{\textcolor[rgb]{0.00,0.00,0.00}{#1}}
\newcommand{\blue}[1]{\textcolor[rgb]{0.00,0.00,0.00}{#1}}

\definecolor{wheat1}{rgb}{1.000000,0.905882,0.729412}
\definecolor{LightGray}{rgb}{0.827451,0.827451,0.827451}

\newcolumntype{a}{>{\columncolor{wheat1}}l}

\definecolor{mygreen}{rgb}{0,0.6,0}
\definecolor{mygray}{rgb}{0.5,0.5,0.5}
\definecolor{mymauve}{rgb}{0.58,0,0.82}
\definecolor{darkblue}{rgb}{0.0,0.0,0.6}
\definecolor{maroon}{RGB}{102, 0, 0}
\definecolor{Maroon}{cmyk}{0,0.87,0.68,0.32}
\definecolor{darkred}{RGB}{139, 0, 0}
\definecolor{forestgreen}{RGB}{34, 139, 34}

\lstdefinelanguage{XML}
{
  basicstyle=\ttfamily\small,   
  morestring=[b]",
  moredelim=[s][\color{darkblue}]{<}{\ },
  moredelim=[s][\color{darkblue}]{</}{>},
  moredelim=[l][\color{darkblue}]{/>},
  moredelim=[l][\color{darkblue}]{>},
  morecomment=[s]{<?}{?>},
  morecomment=[s]{<!--}{-->},
  stringstyle=\color{darkred},
  identifierstyle=\color{mymauve}
}

\lstdefinestyle{customJava}{
  breaklines=true,
  keepspaces=true,
  frame=single,
  language=Java,
  showstringspaces=false,
  basicstyle=\footnotesize\ttfamily,
  keywordstyle=\color{blue},
  otherkeywords={+, getIntent},
  numbers=left,
  numbersep=5pt,
  numberstyle=\scriptsize\color{black},
  rulecolor=\color{black},
  stepnumber=1,
  tabsize=2,
  commentstyle=\itshape\color{green!40!black},
  stringstyle=\color{orange},
  emph=[1]  
  {
        do,
        try,
        new,
        catch,
        while,
        SecProvider,
        SecReceiver,
        SecService,
        SecActivity,
        SecSink,
  },
  emphstyle=[1]{\color{darkred}},
  emph=[2]  
  {
        @Override,
  },
  emphstyle=[2]{\color{purple!40!black}},
  belowskip=-1em, 
}

\newif\ifANNOYMIZE
\ANNOYMIZEfalse

\newif\ifACM
\ACMfalse 

\ifACM
\fi

\ifACM
\newcommand{\myfig}{Figure\xspace}
\else
\newcommand{\myfig}{Fig.\xspace}
\fi

\ifACM
\newcommand{\mysec}{\S}
\else
\newcommand{\mysec}{Sec.\xspace}
\fi

\newcommand{\name}{BackDroid\xspace}
\newcommand{\backdroid}{BackDroid\xspace}
\newcommand{\dexdump}{\texttt{dexdump}\xspace}

\begin{document}
%
\title{\name: On-the-fly Static Dataflow Analysis via Bytecode Search for Targeted Security Vetting of Android Apps}
\title{\name: On-the-fly Static Analysis via Bytecode Search for Targeted Vetting of Android Apps}
\title{\name: Targeted Static Vetting of Android Apps via On-the-fly Bytecode Search}
\title{\name: Targeted Inter-procedural Analysis of Modern Android Apps via On-the-fly Bytecode Search}
\title{Following Devil's Footprints: Cross-Platform Analysis of Potentially Harmful Libraries on Android and iOS}
\title{Finding Unknown Malice in 10 Seconds: Mass Vetting for New Threats at the Google-Play Scale}
\title{When Program Analysis Meets Bytecode Search: Targeted and Efficient Inter-procedural Analysis of Modern Android Apps in \name}


\author{
\ifANNOYMIZE
\IEEEauthorblockN{Anonymous Submission}
\else
\IEEEauthorblockN{Daoyuan Wu$^1$, Debin Gao$^2$, Robert H. Deng$^2$, and Rocky K. C. Chang$^3$}
\IEEEauthorblockA{$^1$The Chinese University of Hong Kong\\
$^2$Singapore Management University\\
$^3$The Hong Kong Polytechnic University\\
$^{\textrm{\Letter}}$ Contact: dywu@ie.cuhk.edu.hk}
\fi
}

\maketitle

\begin{abstract}



Widely-used Android static program analysis tools, e.g., Amandroid and FlowDroid, perform the \textit{whole-app} inter-procedural analysis that is comprehensive but fundamentally difficult to handle modern (large) apps.
The average app size has increased three to four times over five years.
In this paper, we explore a new paradigm of \textit{targeted} inter-procedural analysis that can skip irrelevant code and focus only on the flows of security-sensitive sink APIs.
To this end, we propose a technique called \textit{on-the-fly bytecode search}, which searches the \red{disassembled} app bytecode text just in time when a caller needs to be located.
In this way, it guides targeted (and backward) inter-procedural analysis step by step until reaching entry points, \blue{without relying on a whole-app graph}.
Such search-based inter-procedural analysis, however, is challenging due to Java polymorphism, callbacks, \red{asynchronous flows}, static initializers, and inter-component communication in Android apps.
We overcome these unique obstacles in our context by proposing a set of bytecode search mechanisms that utilize flexible searches and forward object taint analysis.
Atop of this new inter-procedural analysis, we further adjust the traditional backward slicing and \red{forward constant propagation} to provide the \red{complete} dataflow tracking of \red{sink API calls}.
We have implemented a prototype called \backdroid and compared it with Amandroid in analyzing \red{3,178} modern popular apps for crypto and SSL misconfigurations.
The evaluation shows that for such sink-based problems, \backdroid is 37 times faster (2.13 v.s. 78.15 minutes) and has no timed-out failure (v.s. 35\% in Amandroid), while maintaining close or even better detection effectiveness.

\end{abstract}


%

\section{Introduction}
\label{sec:BKintro}

Static analysis is a common program analysis technique extensively used in the software security field.
Among existing Android static analysis tools, Amandroid~\cite{Amandroid14, Amandroid18} and FlowDroid~\cite{FlowDroid14, FlowDroid16Thesis} are the two most advanced \red{and widely used} ones, with the state-of-the-art dataflow analysis capability.
Both perform the \textit{whole-app} inter-procedural analysis that starts from all entry points and ends in all reachable code nodes.
Such analysis is comprehensive \red{but ignores specific analysis requirements, and often comes} at the cost of huge overheads.
For example, a dataflow mining study~\cite{MudFlow15} based on FlowDroid used a compute server with 730 GB of RAM and 64 CPU cores.
Even with such a powerful configuration, ``\textit{the server sometimes used all its memory, running on all cores for more than 24 hours to analyze one single Android app}''~\cite{MudFlow15}.

As a result, small apps were often used in prior studies.
For example, AppContext~\cite{AppContext15} selected apps under 5MB for analysis because they found that not enough apps could be successfully analyzed by its underlying FlowDroid tool.
Although HSOMiner~\cite{HSOMiner17} increased the size of the analyzed apps, the average app size is still only 8.4MB.
Even for these small apps, AppContext timed out for 16.1\% of the 1,002 apps tested and HSOMiner similarly failed in 8.4\% of 3,000 apps, causing a relatively high failure rate.
\red{Hence, this is not only a performance issue, but also the detection burden.}
\blue{Additionally}, third-party libraries were often ignored.
For example, Amandroid by default skipped the analysis of 139 popular libraries, such as AdMob, Flurry, and Facebook.

However, the size of modern apps keeps on increasing.
According to our measurement, \red{the average and median size of popular apps on Google Play have increased three and four times, respectively, over a period of five years.}
With these modern apps, we re-evaluate the cost of generating a relatively precise whole-app call graph using the latest FlowDroid, and find that even this task alone (\red{i.e., without conducting the subsequently more expensive dataflow analysis}) could be \red{sometimes} expensive --- 24\% apps failed even after running for 5 hours each. 
Hence, a scalable Android static analysis is needed \red{to keep pace with the upscaling trend in modern apps}. 
Fortunately, security studies are usually interested only in a small portion of code that involves the flows of security-sensitive sink APIs.
For example, Android malware detection~\cite{AndroidMalware12} is mostly interested in the sink APIs that can introduce security harms (e.g., \texttt{sendTextMessage()}), and vulnerability analysis often spots particular patterns of the code~\cite{Study11}. 
Therefore, it is possible for security-oriented tools to perform a targeted analysis from the selected sinks.


In this paper, we explore a new paradigm of \textit{targeted} (v.s. the traditional whole-app) inter-procedural analysis that can skip irrelevant code and focus only on the flows of security-sensitive sink APIs.
To achieve this goal, we propose a technique called \textit{on-the-fly bytecode search}, which searches the \red{disassembled} app bytecode text just in time when a caller needs to be located \red{so that it can guide targeted (and backward) inter-procedural analysis step by step until reaching entry points}.
We \red{combine this technique with the traditional program analysis} and develop a static analysis tool called \backdroid, for the efficient and effective targeted security vetting of \textit{modern} Android apps.
Since the whole-app analysis is no longer needed in \backdroid, the required CPU and memory resources \blue{are controllable} regardless of app size.
Such a design, however, requires us to solve several unique challenges.
Specifically, it is challenging to perform effective bytecode search over Java polymorphism (e.g., super classes and interfaces), callbacks (e.g., \path{onClick()}), \red{asynchronous flows} (e.g., \path{AsyncTask.execute()}), static initializers (i.e., static \texttt{<clinit>()} methods), and ICC (inter-component communication) in Android apps.
\red{Note that these obstacles are different \blue{from when} they appeared previously~\cite{Amandroid14, FlowDroid14}, where the challenge was to determine object types instead of hindering the searches.}


To overcome those obstacles in our context, we propose a set of bytecode search mechanisms that utilize flexible searches and \red{object taint analysis}.
First, we present \blue{a method signature based} search that constructs appropriate search signatures to directly locate callers for static, private, and constructor callee methods.
\red{This basic search also works well for child classes by just launching one more signature search with the child class.}
Second, for complex situations with super classes, interfaces, callbacks, and asynchronous flows, we propose an advanced search mechanism, \red{because directly searching callers' signatures in these situations would hit nothing}.
\blue{Instead}, we first search the callee \blue{class's} object constructor(s) that can be accurately located, and from those constructors, we perform \red{forward object taint analysis} until reaching caller sites.
\red{Third, we further propose several special search mechanisms, including}
(i) a recursive search to determine the reachability of static initializers;
(ii) a two-time ICC search that first searches for both ICC calls and parameters, and then merges their search results; and
(iii) an on-demand search for Android lifecycle handlers, e.g., \texttt{onStart()} and \texttt{onResume()} methods.

Atop our new paradigm of inter-procedural analysis, we further adjust the traditional backward slicing and \red{forward constant propagation} to provide the dataflow tracking of \red{sink API calls}.
Specifically, we first generate a structure called \textit{self-contained slicing graph} (SSG) to record all the slicing information and inter-procedural relationships during the search-based backtracking. 
To faithfully construct \blue{an} SSG, we not only reserve the raw typed bytecode statements but also taint across fields, arrays, and contained methods, as well as add off-path static initializers on demand.
On top of the generated SSGs, we then conduct \red{forward constant and points-to propagation} over each SSG node to propagate and calculate \red{the complete dataflow representation of target sink API parameters}.



To evaluate \backdroid's efficiency and efficacy, we compare it with the state-of-the-art Amandroid~\cite{Amandroid14, Amandroid18} tool in analyzing modern apps for crypto and SSL misconfigurations, two \red{common and serious} sink-based problems that were \red{also} recently tested by Amandroid in~\cite{Amandroid18}. 
Specifically, we first select a set of 3,178 modern apps that have at least one million installs each and also were updated \red{in recent years} \blue{as our basic dataset}. 
We then \blue{pre-search} them to obtain 144 apps with all the relevant sink APIs so that Amandroid would not waste its analysis \red{even without the bytecode search capability; otherwise, the performance gap between Amandroid and \backdroid could be larger}.
\red{The average and median size of these apps are 41.5MB and 36.2MB, respectively.}
We use a default parameter configuration of Amandroid and run both tools on a machine with 8-core Intel i7-4790 CPU and 16GB of physical memory, a memory configuration often used in many previous studies (e.g.,~\cite{AppSealer14, AppContext15, DomainSocket16, HSOMiner17}).
Moreover, we give Amandroid sufficient running time with a large timeout of five hours \blue{(or 300 minutes)} per app \red{(the timeout setting explicitly reported in prior studies~\cite{CryptoLint13, StubDroid16, HSOMiner17, AppContext15} was 30, 60, and 80 minutes)}. 

Our evaluation shows that \backdroid achieves much better performance while its detection effectiveness is close to, or even better than, that of Amandroid.
First, \backdroid's overall performance is 37 times faster than that in Amandroid, requiring only 2.13min (or minutes) for the median analysis time while that in Amandroid is 78.15min.
\red{Indeed, \backdroid can quickly analyze one third of the apps within one minute each (v.s. 0\% in Amandroid), and 77\% apps can be finished within 10 minutes each (v.s. 17\% in Amandroid).}
Moreover, \backdroid \red{has no timed-out failure and only three apps exceeding 30min.}
In contrast, the \red{timed-out failure} rate in Amandroid is as high as 35\%.
On the other hand, \backdroid still maintains close detection effectiveness for the 30 vulnerable apps detected by Amandroid: \backdroid uncovered 22 of 24 true positives and avoided 6 false positives.
Furthermore, \backdroid discovered 54 additional apps with potentially insecure ECB and SSL issues.
One half were due to the \red{timed-out failures}, but the rest were caused by the skipped libraries, unrobust handling of \red{asynchronous flows}/callbacks, and occasional errors in Amandroid's whole-app analysis.

The remainder of this paper is organized as follows.
We first introduce background and motivation in \mysec\ref{sec:motivate}, and give an overview of \backdroid in \mysec\ref{sec:BKoverview}.
We then present \backdroid's on-the-fly bytecode search technique in \mysec\ref{sec:BKsearch}, followed by the adjusted dataflow analysis in \mysec\ref{sec:BKimplement}.
In \mysec\ref{sec:BKevaluate}, we evaluate \backdroid and compare it with Amandroid.
After discussing the limitations and related works in \mysec\ref{sec:BKdiscuss} and \mysec\ref{sec:related}, \blue{respectively,} we conclude this paper in \mysec\ref{sec:BKconclude}.

\textbf{Availability:} To allow more usages of \backdroid and facilitate future research, we will release the source code of \backdroid on GitHub and maintain its further development.

\section{\red{Background and Motivation}}
\label{sec:motivate}

\subsection{Android Static Analysis Background}
\label{sec:background}

Different from the classical program analysis (e.g., for Java and C/C++) that handles only one entry point, static analysis of Android apps needs to consider many entry points that are implicitly called by the Android framework. 
These entry points are the lifecycle handler methods (e.g., \texttt{onCreate()}) of different components registered in an app configuration file called \texttt{AndroidManifest.xml} (or manifest thereafter).
An important task of Android static security analysis is to test the reachability from entry points to security-sensitive sink API calls.
\red{Since such control-flow reachability usually involves multiple Java methods (or procedures, formally), an \textit{inter}-procedural, instead of \textit{intra}-procedural, analysis is a must in Android static analysis.
Otherwise, we cannot determine whether a sink API call is valid, i.e., not the dead code or from uninvoked libraries (both are common in Android apps).}

Existing Android static tools need to launch the \textit{whole-app} analysis for inter-procedural analysis.
Specifically, before they can perform any inter-procedural dataflow taint~\cite{AndroidLeaks12} or backward slicing~\cite{PiOS11} analysis, they all generate certain kinds of whole-app graphs.
Some are constructed systematically, such as a points-to graph for all objects in Amandroid~\cite{Amandroid14}, a lifecycle-aware call graph in FlowDroid~\cite{FlowDroid14}, a system dependency graph in R-Droid~\cite{RDroid16}, and a context-sensitive call graph in CHEX~\cite{CHEX12}, DroidSafe~\cite{DroidSafe15}, and IntelliDroid~\cite{IntelliDroid16}.
Among them, Amandroid's whole-app graph, generated along with the dataflow analysis, is the most precise one.
Others are built in an ad-hoc manner and are thus less accurate, e.g., a class hierarchy based control flow graph~\cite{Woodpecker12, CryptoLint13, CredMiner15}, a method signature based call graph~\cite{SliceDroid13}, and an \textit{intra}-procedural type-inference based call graph~\cite{BrokenFinger18}.
\red{To the best of our knowledge, we are the first \textit{not} to perform whole-app analysis but still enable effective inter-procedural analysis.}

To build Android static analyses, some common technical tools are often used.
Notably, the Soot~\cite{Soot11} and WALA~\cite{WALAHomepage} analysis frameworks have been used in many prior works.
They provide the underlying intermediate representation (IR), call graph generation, and some basic support of data flow analysis.  
\blue{In \backdroid, we also leverage Soot's Shimple IR (an IR in the Static Single Assignment form) to build our own dataflow analysis but use bytecode search, instead of Soot's call graph, to support the inter-procedural analysis.} 

\subsection{The Upscaling Trend of \red{Modern} App Sizes}
\label{sec:appsize}



Both Amandroid and FlowDroid were initially proposed in 2014.
Although they are still being maintained and improved over these years, handling large apps was not considered as a design objective at the first place.
However, as we will show in this subsection, modern popular apps have increased their sizes dramatically over a period of five years from 2014 to 2018.
This is not too surprising, considering that 1,448 of the top 10,713 apps studied in 2014~\cite{FreshApps15} had already been updated on a bi-weekly basis or even more frequently.
\red{Note that although app size increase is not fully due to code change (also related to user interface XML code and resource change), it is the most common metric used to describe an app dataset, and allows a comparison between our and prior datasets (e.g.,~\cite{AppContext15, HSOMiner17}).}

To measure the changes in the app sizes, we first obtain a set of modern apps.
Specifically, we collected a set of 22,687 popular apps on Google Play in November 2018 by correlating the AndroZoo repository~\cite{AndroZoo16} with the top app lists available on \url{https://www.androidrank.org}.
Each app in this set had at least one million installs on Google Play.
We then recorded the app sizes and DEX file dates (if any) in our dataset.

\begin{table}[t!]
\caption{A summary of average and median app sizes over a period of five years from 2014 to 2018.}
\label{tab:appsizechange}
\begin{adjustbox}{center}
\scalebox{1.2}{
\begin{tabu}{|c|c|c|c|} 

\hline
\rowfont{\bfseries}
\rowcolor{LightGray}
Year & Average Size & Median Size   & \# Samples \\
\hline
\hline

2014 & 13.8MB       & 8.4MB         & 2,840 \\
2015 & 18.8MB       & 12.4MB        & 1,375 \\
2016 & 21.6MB       & 16.2MB        & 3,510 \\
2017 & 32.9MB       & 30.0MB        & 1,706 \\
2018 & 42.6MB       & 38.0MB        & 3,178 \\
\hline

\end{tabu}
}
\end{adjustbox}
\end{table}

Table~\ref{tab:appsizechange} summaries \blue{the average and median app sizes} over a period of five years from 2014 to 2018.
We can see that in 2014, the average and median app size is only 13.8MB and 8.4MB, respectively.
This number almost doubles in 2016, with an average size of 21.6MB and a median size of 16.2MB.
It further doubles after two years, with an average app size of 42.6MB in 2018.
This clearly shows that modern apps have dramatically increased their app sizes, \red{and they are expected to further enlarge as more functionalities are added}.



\subsection{The Cost of Generating a Whole-app Call Graph for Modern Apps}
\label{sec:callgraph}

With modern apps, we now re-evaluate the cost of generating a relatively precise whole-app call graph.
Specifically, we measure the execution time required by FlowDroid 2.7.1\footnote{It is the latest release at the time of our submission in May 2020. Note that \red{this version of FlowDroid has continued improving the performance on top of the most recently published one} in 2016~\cite{FlowDroid16Thesis, FlowDroidGithub} for over two years.} to analyze a set of 144 modern apps with the average size of 41.5MB, under the same hardware configuration for our experiments of \backdroid and Amandroid in \mysec\ref{sec:BKevaluate}.  
We choose FlowDroid because it decouples the logic of call graph generation and dataflow taint analysis (whereas Amandroid cannot), which allows us to measure the cost of generating call graphs only.
To increase the precision of FlowDroid's call graph generation, we use the context-sensitive geomPTA~\cite{geomPTA} call graph algorithm, instead of the context-insensitive SPARK algorithm~\cite{FlowDroidCallGraph}.
However, we do not launch IccTA~\cite{IccTA15} to further transform the generated call graphs with inter-component edges.
This sacrifices certain accuracy but keeps stability.

\begin{figure}[t!]
\begin{adjustbox}{center}
\includegraphics[width=0.4\textwidth]{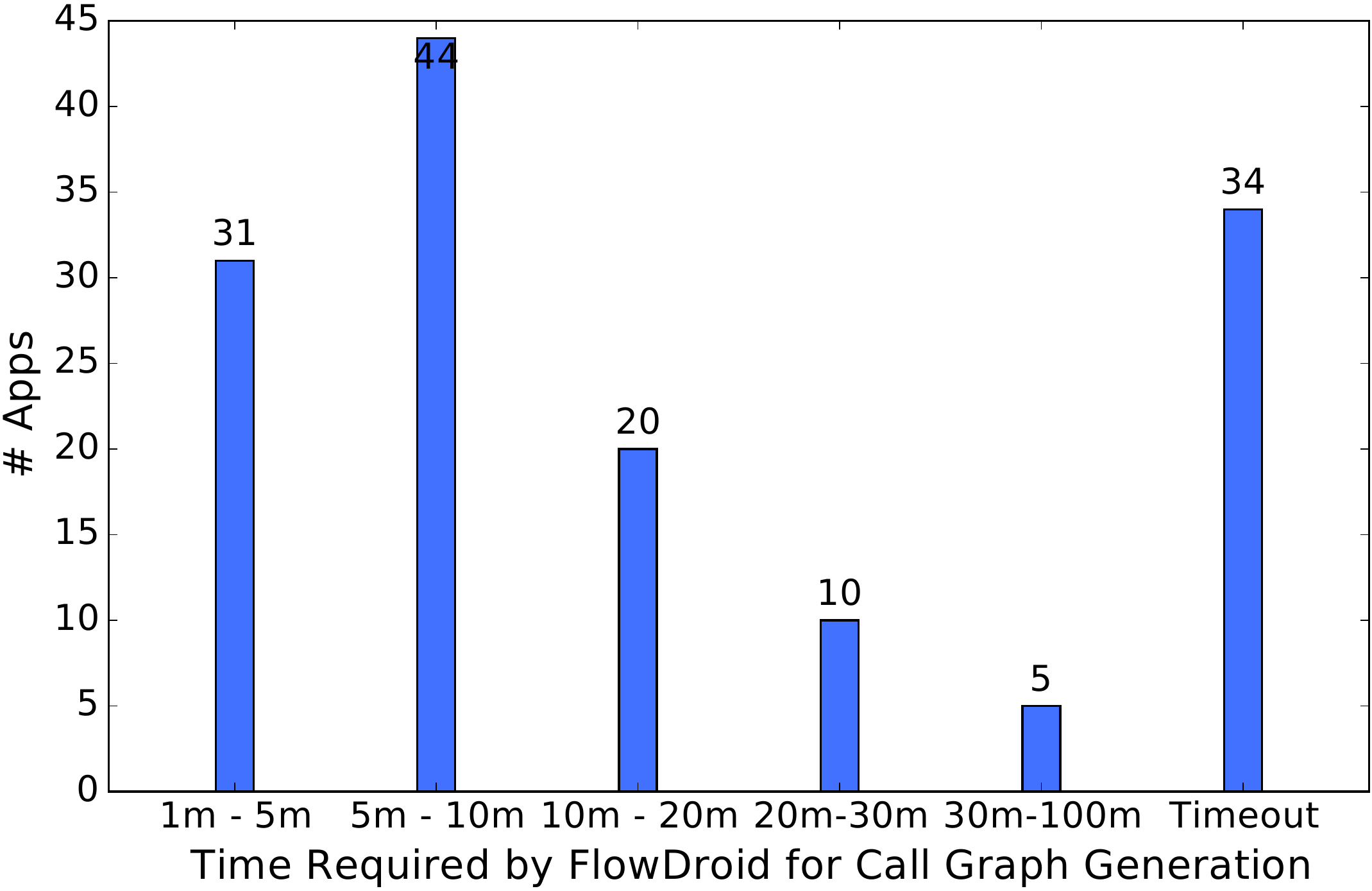}
\end{adjustbox}
\caption{\blue{FlowDroid's call graph generation time for a set of 144 modern apps (under a timeout of 5 hours each).}}
\label{fig:flowTimeBar}
\end{figure}

\red{
We find that even generating a whole-app call graph alone (without performing the subsequently more expensive dataflow analysis) could be sometimes expensive.
Figure~\ref{fig:flowTimeBar} presents FlowDroid's call graph generation time for a set of 144 modern apps.
Although the call graphs of 31 (21.5\%) apps could be quickly generated within five minutes, the median time of call graph generation in FlowDroid is still around 10 minutes (9.76min) per app.
Considering the total analysis time required by \backdroid for the same set of apps is only 2.13min (see \mysec\ref{sec:BKevaluate}), such a call graph generation is already 4.58 times slower.
Besides the performance concern, the detection burden caused by timed-out failures is much more serious.
We can see that as high as 24\% apps failed even after running for 5 hours each, causing no any analysis result could be outputted for these 34 modern apps among the total 144 apps.
}

\section{Overview and Challenges}
\label{sec:BKoverview}

\begin{figure*}[t!]
\begin{adjustbox}{center}
\includegraphics[width=1\textwidth]{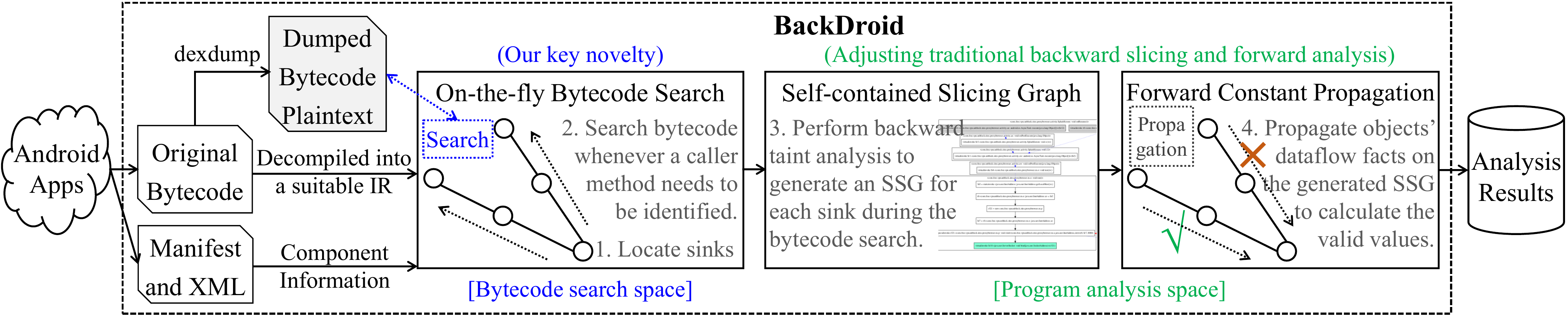}
\end{adjustbox}
\caption{A high-level overview of \backdroid.}
\label{fig:backdroidoverview}
\end{figure*}

Motivated by the incapability of \textit{whole-app} inter-procedural analysis to analyze modern Android apps, we explore a new paradigm of \textit{targeted} inter-procedural analysis that can skip irrelevant code and focus only on the flows of security-sensitive sink APIs.
This new paradigm is enabled by a technique called \textit{on-the-fly bytecode search}, which searches the \red{disassembled} app bytecode text just in time to guide targeted (and backward) inter-procedural analysis until reaching entry points.
We implement this technique into a tool called \backdroid, and further adjust the traditional backward slicing and forward constant propagation to provide the dataflow tracking of targeted sink APIs.
\myfig~\ref{fig:backdroidoverview} presents a high-level overview of \backdroid, which works in the following four major steps:

\begin{compactenum}
\item \textit{Preprocessing with \red{disassembling} bytecode into plaintext.}
    Given an input of any Android app(s), \backdroid first extracts original bytecode and manifest files.
    After that, it not only transforms bytecode into a suitable intermediate representation (IR) as in typical Android analysis tools, but also employs \dexdump~\cite{dexdump} to \red{disassemble} (merged, if multidex~\cite{multidex} is used) bytecode to a plaintext.
 
  \item \textit{Bytecode search for targeted inter-procedural analysis.}
    With the \red{disassembled} bytecode plaintext, \backdroid immediately locates the target sink API calls by performing a text search of bytecode plaintext \red{(in the bytecode search space)} and initiates the analysis from there \red{(in the program analysis space)}.
    To support the inter-procedural analysis with no call graph, \backdroid performs on-the-fly bytecode search to identify caller methods \blue{whenever needed}.

  \item \textit{Performing the adjusted backward slicing to generate SSG.}
    To also provide the dataflow tracking atop our search-based inter-procedural analysis, \backdroid performs the traditional backward slicing but adjusts it into our new context.
    Specifically, during the inter-procedural backtracking, \backdroid generates a self-contained slicing graph (SSG) for each sink API call to record all the slicing information and inter-procedural relationships \red{resolved by search}.

  \item \textit{Forward analysis over SSG to propagate and calculate dataflow.}
    On top of the generated SSGs, \backdroid launches the classical forward constant propagation~\cite{ConstantPropagation91} to propagate and calculate dataflow facts from entry points to sink APIs.
    \red{It also propagates object points-to~\cite{JavaPointsTo03} information to remove potential ambiguity.} 
    Eventually, with the support of SSG, \backdroid is able to output the \textit{complete} dataflow representation (either a constant or an expression) of target sink API parameters.
\end{compactenum}

\textbf{Challenges.}
Given that \backdroid is the first inter-procedural analysis tool without relying on a whole-app graph, its major novelty and biggest challenge is how to perform on-the-fly bytecode search to locate caller methods in a backward manner.
This is difficult because of Java polymorphism (e.g., super classes and interfaces), callbacks, \red{asynchronous Java/Android flows}, static initializers, and inter-component communication, all of which make a basic signature based search infeasible.
We will present our core search technique in \mysec\ref{sec:BKsearch}.
Another challenge is how to adjust the traditional backward slicing and forward analysis into our new paradigm of targeted inter-procedural analysis.
We will present its implementation challenges and our corresponding solutions in \mysec\ref{sec:BKimplement}.

\section{On-the-fly Bytecode Search}
\label{sec:BKsearch}

In this section, we present our novel bytecode search technique to locate caller methods on the fly, which is the key to \backdroid's targeted inter-procedural analysis.
We first present the basic signature based search in \mysec\ref{sec:basicSearch} and an advanced search mechanism with forward object taint analysis in \mysec\ref{sec:advancedSearch}.
We then elaborate three special search mechanisms from \mysec\ref{sec:searchStaticInit} to \mysec\ref{sec:searchLifecycle}, on how to effectively search over static initializers, Android ICC (inter-component communication), and Android lifecycle handlers (e.g., \texttt{onStart()} and \texttt{onResume()} methods).
Finally in \mysec\ref{sec:enhancements}, we describe several implementation enhancements to the performance of our bytecode search.


\subsection{Basic Search by Constructing Appropriate Search Signature(s)}
\label{sec:basicSearch}

\begin{figure*}[t!]
\begin{adjustbox}{center}
\includegraphics[width=0.85\textwidth]{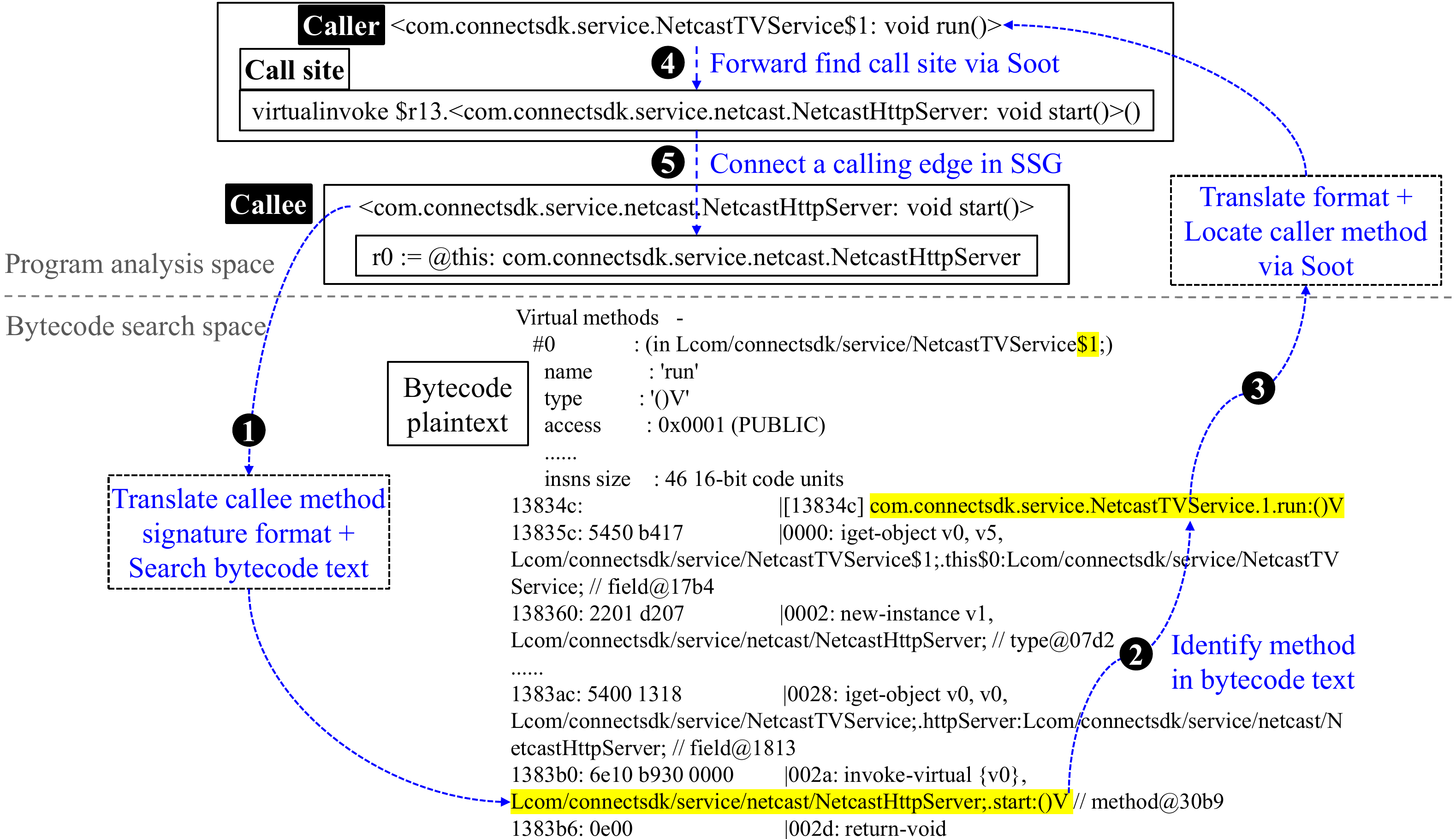}
\end{adjustbox}
\caption{Illustrating \backdroid's basic bytecode search process using a signature based search example.}
\label{fig:basicSearch}
\end{figure*}

To illustrate our search process, we use a real popular app, LG TV Plus\footnote{\url{https://play.google.com/store/apps/details?id=com.lge.app1}}, which has over 10 million installs on Google Play, as a running example.
In \myfig~\ref{fig:basicSearch}, we have already used initial bytecode search to find a target method (i.e., the one with a sink API call), \path{<com.connectsdk.service.netcast.NetcastHttpServer:} \path{void} \path{start()>}.
For inter-procedural analysis, our next step is to uncover its caller method (i.e., \path{<com.connectsdk.service.NetcastTVService}\texttt{\$}\path{1:} \path{void} \path{run()>}) and its call site (i.e., statement \path{virtualinvoke} \texttt{\$}\path{r13.<com.connectsdk.service.netcast.NetcastHttpServer:} \path{void} \path{start()>()}).
As we will explain, searching its caller can be done directly with the following (method) signature based search, because the target callee method here is a \texttt{private} Java method.

\textbf{The basic signature based search.}
As illustrated in \myfig~\ref{fig:basicSearch}, we conduct the signature based search in five steps, which are across not only the bytecode search space but also the program analysis space (with Soot).
Given a callee method, we first translate its method signature from Soot's IR format to \texttt{dexdump}'s bytecode format to facilitate the search.
With the transformed method signature, we then search the entire bytecode plaintext to locate all its invocation(s), as highlighted in the bottom of \myfig~\ref{fig:basicSearch}.
In the second step, we identify the corresponding method that contains the invocation found in the bytecode plaintext.
Here it is \path{com.connectsdk.service.NetcastTVService.}\texttt{\$}\path{1.run:()V}, where an inner class needs to add back the symbol ``\texttt{\$}''.
With this caller method signature (in bytecode format), we perform another format translation in the third step, and locate its Java method body via the program analysis in Soot.
Next, we conduct a quick forward analysis via Soot to find the actual call site in the caller method body.
With all these steps done, we finally connect a edge from the caller (site) to the callee method in SSG (self-contained slicing graph).
Note that SSG is generated during the process of bytecode search and backward slicing.
We will explain the details of SSG and its generation in \mysec\ref{sec:generateBSG}.

After understanding this search process, we now turn to an important question not answered yet: which kinds of (callee) methods are suitable for the signature based search.
We call such methods \textit{signature methods}.
Typical signature methods include static methods\footnote{It is worth noting that although the static \texttt{<clinit>} method of a class is a signature method, it has to use a special search instead of the basic signature based search, as we will explain in \mysec\ref{sec:searchStaticInit}.} (either class or method is marked with the \texttt{static} keyword), private methods (similarly, methods declared with the \texttt{private} keyword), and constructors (e.g., \texttt{<init>} methods of a class).
For some searches over child classes, we can also simply launch the signature based search, as explained next.

\textbf{Searching over a child class.}
Suppose that the \texttt{NetcastHttpServer} class in \myfig~\ref{fig:basicSearch} has a child class called \texttt{ChildServer}, we can still use the signature based search but need to construct appropriate search signatures.
Specifically, it depends on whether \texttt{ChildServer} overloads the callee method \texttt{void start()} or not.
If it is not overloaded, an invocation of the callee method \texttt{start()} may also come from a child class object.
Hence, besides the original signature search, we add one more signature search with the child class, namely \path{Lcom/connectsdk/service/netcast/ChildServer;.start:()V}.
The returned caller(s) might come from both searches, or just one of them, depending on how app developers invoke that particular callee method.
On the other hand, if \texttt{ChildServer} does overload the \texttt{start()} method, we just need to perform only one search with the original callee method signature.
This is because the child class search signature now corresponds to the overloaded child method only.

%

\subsection{Advanced Search with Forward \red{Object Taint} Analysis}
\label{sec:advancedSearch}

\begin{figure*}[t!]
\begin{adjustbox}{center}
\includegraphics[width=0.85\textwidth]{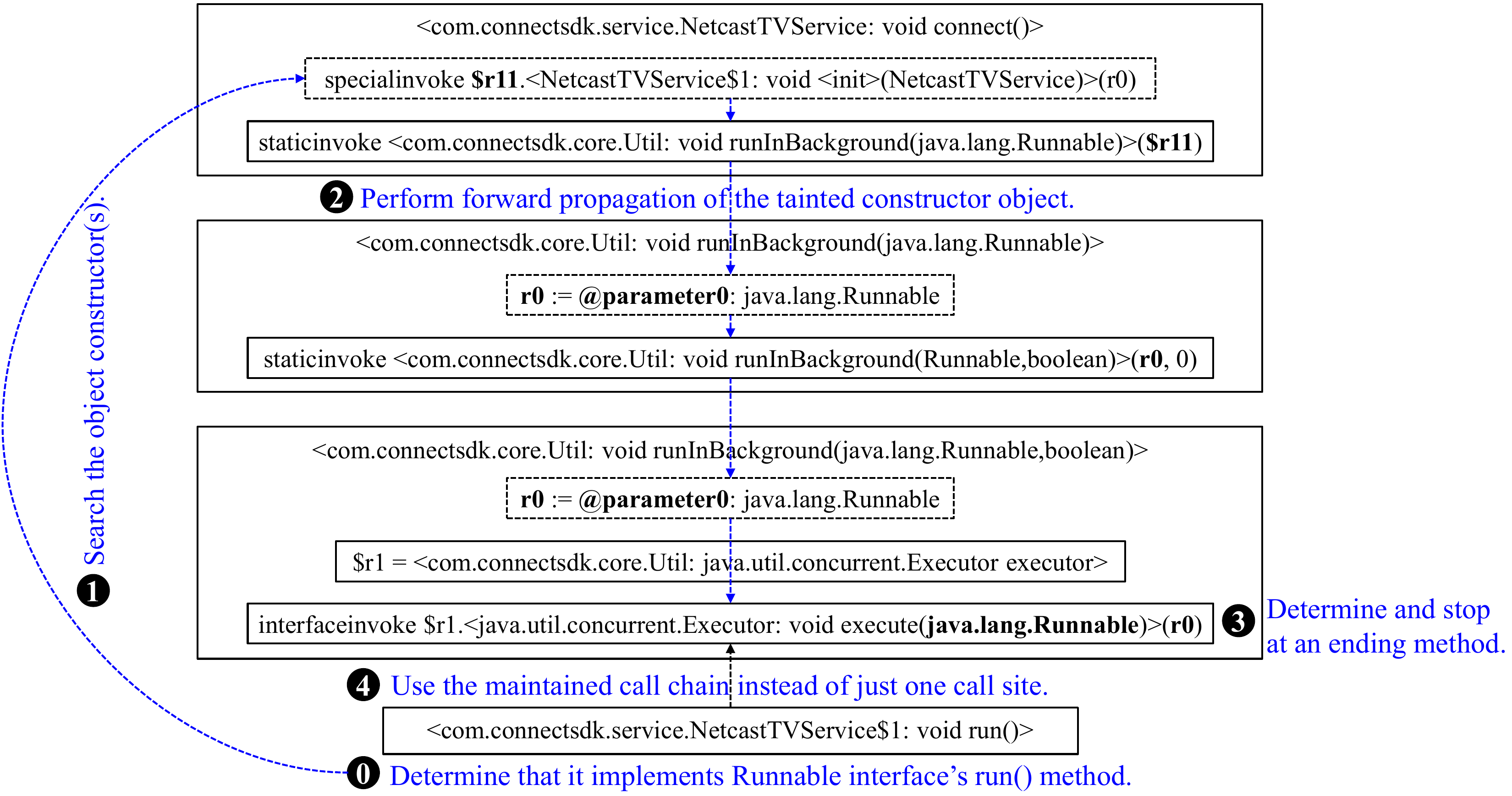}
\end{adjustbox}
\caption{Using advanced search with \red{forward object taint analysis} to uncover a caller chain of an interface method, \texttt{NetcastTVService\$1.run()}. Note that step 1 uses the basic signature search in \mysec\ref{sec:basicSearch}, the process of which is thus skipped.} 
\label{fig:advancedSearch}
\end{figure*}

Although the basic search presented in the last subsection can handle many callee methods in an app bytecode, it is not effective for complex cases with super classes, interfaces, callbacks, and \red{asynchronous Java/Android flows}.
Note that these obstacles are different from the situation when they appeared in the previous research~\cite{Amandroid14, FlowDroid14}, where the challenge was to determine object types instead of hindering the searches.
We first explain the difficulty of searching in the case where \texttt{NetcastHttpServer.start()} in \myfig~\ref{fig:basicSearch} has a super class method called \texttt{SuperServer.start()}.
Under this condition, the original signature search may not reveal any valid callers, because developers may write code in this way: \texttt{SuperServer server = new NetcastHttpServer(); server.start();}.
In this example, the bytecode signature of \path{server.start()} is \path{Lcom/connectsdk/service/netcast/SuperServer;.start:()V}.
Hence, searching with \texttt{NetcastHttpServer}'s method signature would hit nothing.
But we also cannot use super class \texttt{SuperServer}'s signature to launch the search, because it could return callers of the super method itself and other class methods that inherit from \texttt{SuperServer}.
Similarly, if a callee method implements an interface, searching using the interface method signature would not work because an interface method might be implemented by arbitrary classes.
Furthermore, searching over callbacks and \red{asynchronous Java/Android flows} could be even more difficult, because they employ different sub-method signatures for a pair of caller and callee methods.

We design a novel mechanism to accurately handle all these complex searches.
The basic idea is that instead of directly searching for caller methods, we first search the callee class's object constructor(s) that can be accurately located via the signature based search.
\blue{Right} from those object constructors, we then perform \red{forward object taint analysis} until \red{we detect the caller methods with the tainted object propagated into}.
We depict this process in \myfig~\ref{fig:advancedSearch}, using the same LG TV Plus app.
This time the callee method is \path{<com.connectsdk.service.NetcastTVService}\texttt{\$}\path{1:} \path{void} \path{run()>}, which continues the search flow in Fig.~\ref{fig:basicSearch}.
We now present the involved four steps. 

\textbf{Searching for the object constructor.}
After determining a callee method that requires advanced search in step \circled{0}, we first retrieve all its constructors.
In \myfig~\ref{fig:advancedSearch}, the callee class \texttt{NetcastTVService\$1} has only one constructor, \path{void} \path{<init>(com.connectsdk.service.NetcastTVService)}.
We then launch a bytecode search using this signature to \textit{accurately} locate that the constructor is initialized in a method called \path{NetcastTVService:} \path{void} \path{connect()}, as shown in step \circled{1}.

\textbf{Propagating object using taint analysis.}
In step \circled{2}, we perform forward propagation of the located constructor object, i.e., \texttt{\$r11} in \myfig~\ref{fig:advancedSearch}, using taint analysis.
Specifically, an object can be propagated via a definition statement, e.g., \path{r0} \path{:=} \path{@parameter0:} \path{java.lang.Runnable}, via an invoke statement, e.g., \texttt{runInBackground(\$r11)}, or via a return statement.
Therefore, we track only three kinds of statements, namely \texttt{DefinitionStmt}~\cite{SootDefinitionStmt}, \texttt{InvokeStmt}~\cite{SootInvokeStmt}, and \texttt{ReturnStmt}~\cite{SootReturnStmt}.

\textbf{Determining the ending method to stop.}
An important step is to determine at which \textit{ending method} our forward analysis should stop.
This is easy for the case of super class, because we can simply stop at a tainted statement with the same sub-method signature as the callee method.
However, it is difficult for the cases of interface, callback, and \red{asynchronous flow}, because their sub-method signature might be different from that in a callee method.
Some previous works (e.g.,~\cite{Woodpecker12, ECVDetector14, DroidSIFT14}) used the pre-defined domain knowledge to connect those asynchronous flows, e.g., a common example is to connect \texttt{Thread} class'\blue{s} \texttt{start()} and \texttt{run()} methods.
However, in the example of \myfig~\ref{fig:advancedSearch}, it will miss the ending method \texttt{Executor.execute()}. 
We also tried the flow mapping provided by EdgeMiner~\cite{EdgeMiner15} but found that it could contain over-estimated asynchronous flows, such as connecting 2,446 caller methods (e.g., \path{WaveView.<init>()} and \path{Thread.<init>()}) to only a single callee method \texttt{MediaPlayer\$OnErrorListener.onError()}.

To better determine the ending method, we propose a mechanism that does not rely on prior knowledge but leverages interface's class type as an indicator to find a tainted parameter that is directly propagated from the original constructor object.
For example, in \myfig~\ref{fig:advancedSearch}, since the interface class type is \path{java.lang.Runnable}, we determine which on-path Java/Android API call contains a tainted parameter that belongs to this class type.
Our forward analysis thus propagates the original \texttt{\$r11} object along the path highlighted, and eventually reaches at \texttt{r0} in the statement \texttt{Executor.execute(r0)}.
Hence, it is the ending method we need to identify in this example.

%

\textbf{Maintaining and returning a call chain.}
Different from the basic search that returns just one call site, here we need to maintain and return a call chain, i.e., the chain from \path{NetcastTVService.connect()} to \path{Util.runInBackground(Runnable)} and further to \path{Util.runInBackground(Runnable,boolean)}.
Assuming that we just return one call site and one caller method, the further backward search of \path{Util.runInBackground(Runnable,boolean)} may return multiple search results or flows, because this search is now independent.
However, the fact is that only the flow shown in Fig.~\ref{fig:advancedSearch} could eventually trace back to the constructor object.
Therefore, to avoid mis-added flows, we maintain a call chain during the forward taint analysis.

\subsection{Special Search over Static Initializers}
\label{sec:searchStaticInit}

The basic and advanced searches presented in the last two subsections are useful in most scenarios, but they are not for handling some special search challenges.
We thus further propose several special search mechanisms.
In this subsection, we present a recursive search for static initializers.

A static initializer is the static \texttt{<clinit>} method of a class, which may occasionally appear in a call path \backdroid is backtracking.
Resolving its caller methods is a must for determining whether the corresponding call path is reachable from entry points or not. 
However, it is impossible to directly search static initializers' callers, because they are never (explicitly) invoked by any app bytecode.
Instead, \texttt{<clinit>} methods are only \textit{implicitly} executed by Java/Android virtual machine (VM) when the corresponding classes are loaded to the VM.
Hence, we design a special search mechanism to handle these \textit{on-path} static initializers.
Additionally, in \mysec\ref{sec:generateBSG}, we will also show how to add back \textit{off-path} static initializers into our slicing graph for the complete dataflow analysis.

We use another concrete example, the popular Heyzap\footnote{Heyzap was used to connect different advertisement SDKs, and has been acquired by Fiber, a company providing the mobile monetization platform.} advertisement library embedded in many apps, to illustrate our approach. 
In this example, when \backdroid backtracks the \path{setHostnameVerifier()} sink API invoked by \path{com.heyzap.http.MySSLSocketFactory}, the analysis comes to a static initializer of the \path{com.heyzap.internal.APIClient} class.
However, a further search of the \path{APIClient.<clinit>()} initializer method would hit nothing, as we have explained earlier.
Normally, a forward whole-app analysis approach can track into such initializer method whenever it encounters a field or a method of the \texttt{APIClient} class, but here it is unrealistic to launch backward search of all fields and methods of \texttt{APIClient}.
To address this unique challenge in \backdroid, we propose a different approach that performs recursive searches to determine only the control-flow reachability of a targeted \texttt{<clinit>} method.
This is reasonable because \texttt{<clinit>} methods have no dataflow propagation due to no parameter passing.

Formally, for a static initializer \path{SI.<clinit>()}, our recursive search works as follows.
\backdroid first launches a search to find out a set of classes $C = \{c_1, \cdots, c_n\}$ that invoke the \path{SI} class.
It then determines whether any class $c_i$ in this set is an entry component registered in the app manifest.
In the case of Heyzap's \path{APIClient} class, since no $c_i$ is an entry class, \backdroid continues to perform the class search of each $c_i$ to determine whether any of its contained classes is an entry class.
This process repeats until no more new class searched out or entry class identified.
For example, the \path{APIClient} class is invoked by the class \path{com.heyzap.house.model.AdModel}, which is further used by an entry class called \path{com.heyzap.sdk.ads.HeyzapInterstitialActivity}.
Therefore, we consider the initializer \path{APIClient.<clinit>()} is reachable from entry points and the associated call path is also valid.

Through the evaluation in \mysec\ref{sec:BKevaluate}, we have obtained some convincing data in real apps to support our design in this subsection.
Among 37 unique static initializers that are identified by our recursive search \blue{as reachable}, we find that all of them are actually reachable from entry components.

\subsection{Special Search over Android ICC}
\label{sec:searchICC}


In this subsection, we present another special search mechanism to track data flows over Android-specific inter-component communication (ICC), a fundamental code/data collaboration mechanism on Android~\cite{ICCEpicc13, SCLib18}.

Our search is based on the inner working mechanism of Android ICC.
Specifically, ICC is different from typical API calls because it relies on its \texttt{Intent} parameter values to dynamically determine a target callee.
A callee could be \textit{explicitly} specified by setting the target component class (e.g., via \texttt{Intent i = new Intent}\path{(activity,HttpServerService.class);}), or \textit{implicitly} specified by setting an \texttt{Intent} action that will be delivered by the operating system (OS) to the target component.

Based on this observation, we propose a two-time search mechanism to handle ICC.
The basic idea is to launch two searches: one for searching ICC calls (e.g., \path{startService()}), and the other for searching ICC parameters.
For the explicit ICC, the second parameter search directly searches component class names, e.g., \texttt{const-class .*,} \path{Lcom/lge/app1/fota/HttpServerService;}, while for the implicit ICC, we search \texttt{Intent} action names instead.
After performing the searches, we merge the two search results and check whether an ICC call satisfies both.
If such an ICC call exists, it is the caller method we are looking for.


%
%

\subsection{Special Search over Android Lifecycle Handlers}
\label{sec:searchLifecycle}

The last specific search challenge is how to search over Android lifecycle handlers, e.g., the \texttt{onStart()} and \texttt{onResume()} methods in \texttt{Activity} components.
Each lifecycle handler could be an entry function, and they can be executed in multiple orders~\cite{FlowDroid14}.
Our strategy is to first determine whether the dataflow tracking finishes when reaching at a lifecycle handler.
If it does, we have no need to launch further search, since the tracked lifecycle handler is already an entry method.
Otherwise, we conduct a special search that leverages existing domain knowledge~\cite{FlowDroid14} to further track other lifecycle handlers that invoke the callee handler.
Since there are only four kinds of Android components, we can simply use domain knowledge to handle all lifecycle handlers.




\subsection{Implementation Enhancements}
\label{sec:enhancements}

In the course of implementing these on-the-fly bytecode searches, we identify and make several important technical enhancements to guarantee the performance of bytecode search.

\begin{figure}[t!]
\begin{adjustbox}{center}
\includegraphics[width=0.3\textwidth]{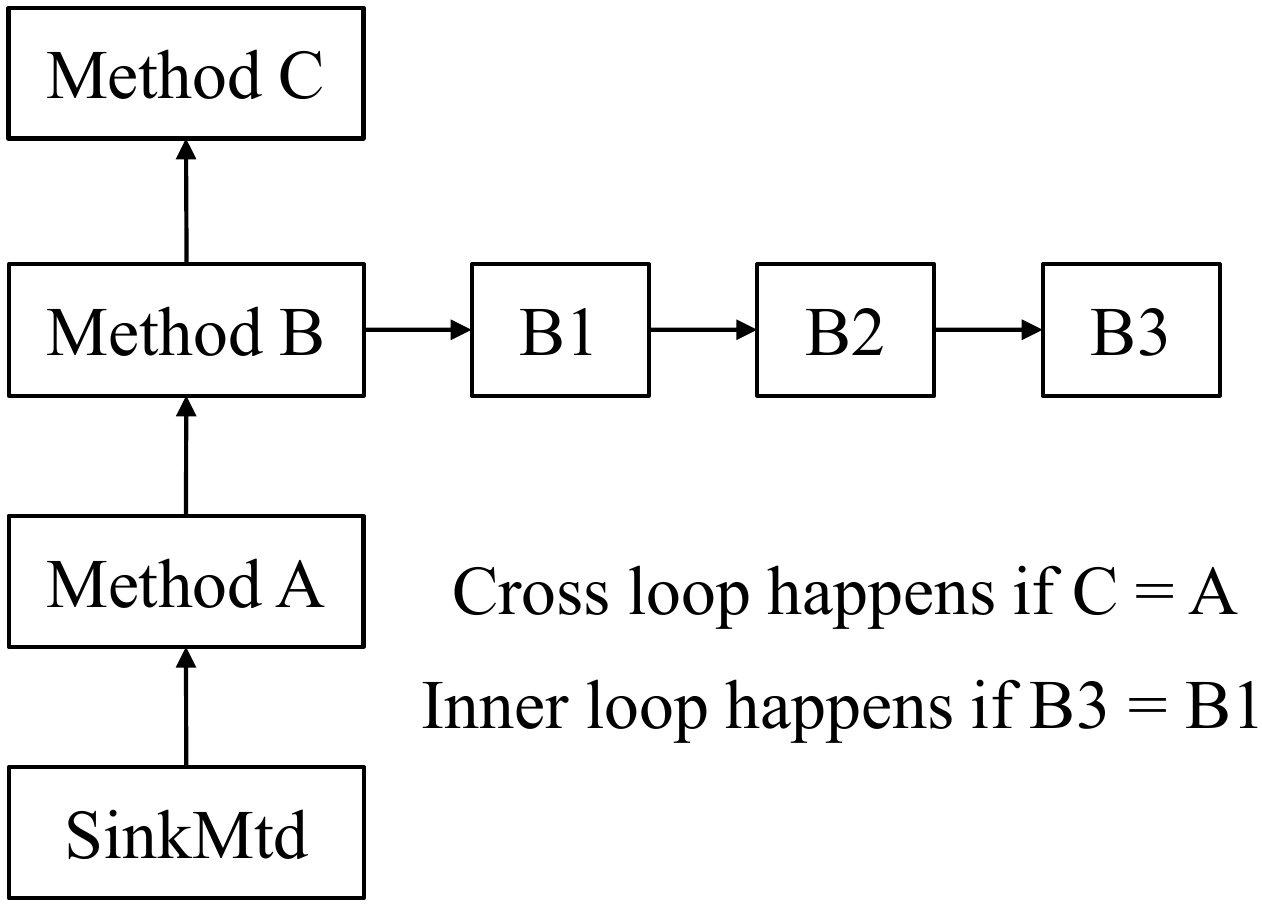}
\end{adjustbox}
\caption{Illustrating two kinds of loops during the backtracking.}
\vspace{-2ex}
\label{fig:deadMtdLoop}
\end{figure}

\textbf{Search caching.}
The first enhancement is to cache different search commands and their corresponding results.
This is necessary because in a valid app analysis, \backdroid will make a number of searches and a portion of them could be executed repeatedly (especially when similar paths are explored across different sinks).
Caching can avoid repeating the same searches.
We perform caching with different granularities, including the caching of invoked class search, caller method search, static or instance field search, and the caching of various raw search commands.
According to our experiment in \mysec\ref{sec:BKevaluate}, the cache rate of our search commands in each app is 23.39\% on average, with the minimum at 2.97\% and the maximum up to 88.95\%.
This data demonstrates the effectiveness of our search caching in \backdroid.

\textbf{Sink API call caching.}
Multiple sink API calls might be located in the same method (e.g., due to \texttt{if-else} statements) that turns out to be unreachable from entry functions.
To avoid re-analyzing such infeasible sink API call, we cache each sink API's callee method signature and its reachability.
If one sink API call is located in a method that has been analyzed and is not reachable, we then do not analyze this sink API call any more. 
Our experiment shows that on average, 13.86\% of sink API calls in each app are cached, with the maximum cache rate at 68.18\%.

\textbf{Method loop detection.}
The third enhancement is to detect the potential dead method loop(s) in our inter-procedural backward search and \red{forward object taint analysis}.
\myfig\ref{fig:deadMtdLoop} illustrates two kinds of method loops during the backward analysis.
First, a cross-method loop could happen during the backward method search.
In \myfig\ref{fig:deadMtdLoop}, our backtracking follows the order of \texttt{SinkMtd --> A --> B --> C}.
However, if the method \texttt{C} is the same as the method \texttt{A}, an infinite search loop happens.
Besides the cross loop, an inner loop could also happen.
As shown in \myfig\ref{fig:deadMtdLoop}, the method \texttt{B} contains a tainted \texttt{invoke} statement and it produces a method call chain from \texttt{B1} to \texttt{B3}.
Suppose that \texttt{B1} and \texttt{B3} are the same, our method tracking would then iterate.
To mitigate the impact of these dead loops, our backtracking will detect them and avoid repeating the analysis.

Besides the backward analysis, our forward object taint analysis in \mysec\ref{sec:advancedSearch} suffers from a similar issue.
To distinguish different loops in \backdroid's running output, we name all the four types of method loops as follows: the \texttt{CrossBackward} and \texttt{InnerBackward} types in the backward scenario, and the \texttt{CrossForward} and \texttt{InnerForward} types in the forward scenario.
By detecting at least one dead method loop per app, we can optimize the path analysis of 60\% apps in our experiment.
Among the four types of search loops, the \texttt{CrossBackward} loop is the most common one.



\section{Adjusting Traditional Backward Slicing and Forward Analysis}
\label{sec:BKimplement}

With the new paradigm of inter-procedural analysis enabled by our on-the-fly bytecode search, we further adjust the traditional backward slicing and forward analysis to provide the dataflow tracking capability~\cite{Amandroid14, FlowDroid14} in our context.
As mentioned earlier in \mysec\ref{sec:BKoverview}, we generate a new slicing structure, self-contained slicing graph (SSG), during the backtracking, and perform forward constant \red{and points-to} propagation over SSG to calculate valid dataflow facts.

\subsection{Generating a Self-contained Slicing Graph (SSG)}
\label{sec:generateBSG}

Since our bytecode search reveals only inter-procedural relationships and we do not have a whole-app graph, we need \red{our own} graph structure to record all the slicing and inter-procedural information during the backtracking.
This graph essentially reflects the partial app paths visited by our on-the-fly analysis, and the information stored in it can be used by forward analysis to recover the complete dataflow representation of target sink API calls.
Below we first describe this graph structure, and then present the implementation challenges we overcame to generate it.

\begin{figure*}[t!]
\vspace{-2ex}
\begin{adjustbox}{center}
\includegraphics[width=1.05\textwidth]{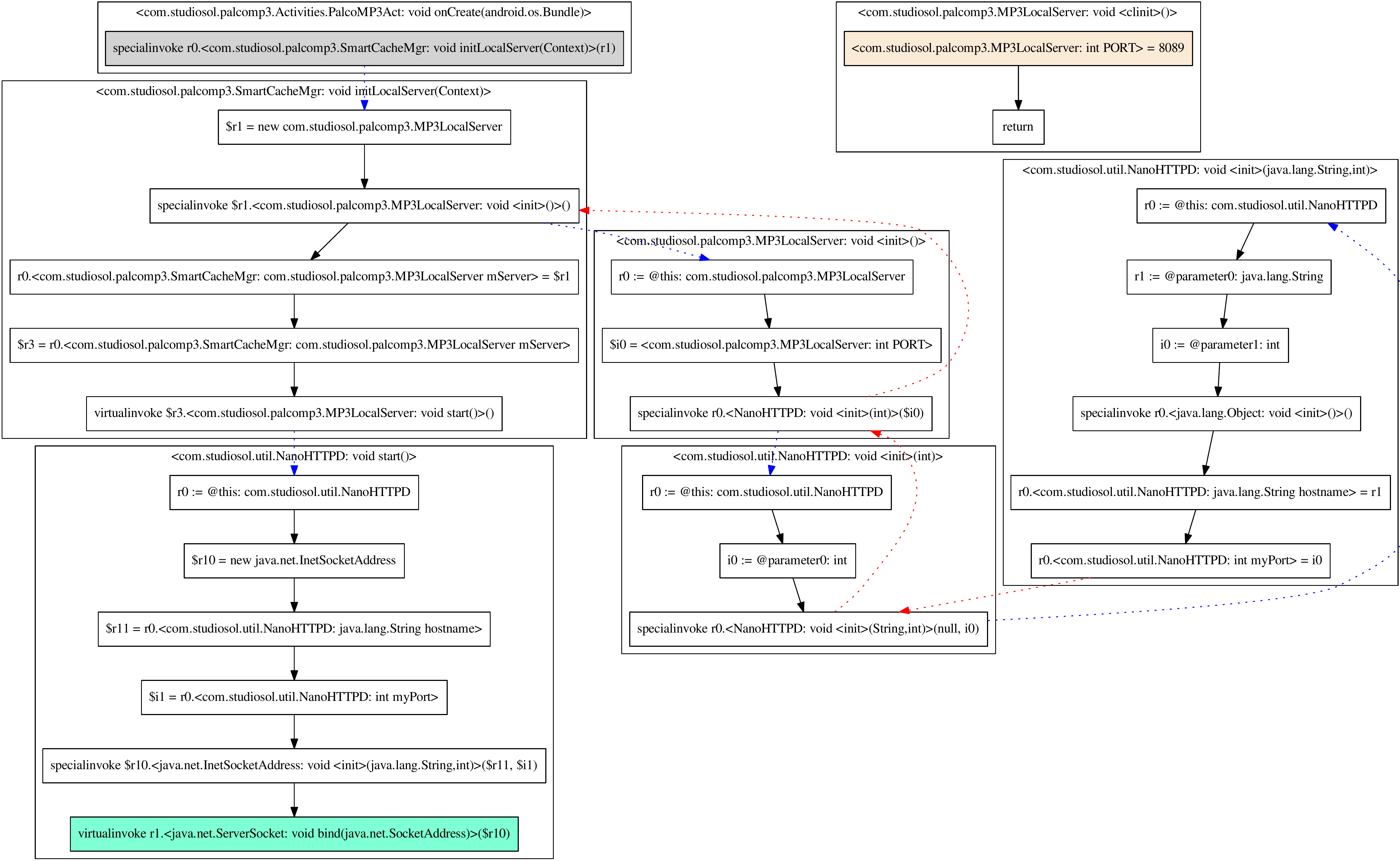}
\end{adjustbox}
\caption{An SSG automatically generated by \backdroid, where the green block is a sink API call and the gray block is an entry point.}
\vspace{-2ex}
\label{fig:exampleBSG}
\end{figure*}

\textbf{Defining a self-contained graph structure to record all the slicing and inter-procedural information.}
We propose a self-contained graph structure called \textit{self-contained slicing graph} (SSG) to cover all the slicing and inter-procedural information generated during our backtracking.
With this structure, we aim to cover the slicing information across different parameters tracked, different paths traced, and different kinds of bytecode instructions, besides recording the inter-procedural relationships uncovered by our bytecode search.
Hence, it is different from the individual path-like slices generated in typical Android slicing tools (e.g.,~\cite{SliceDroid13, RDroid16, CredMiner15}).
We currently design each SSG corresponding to one unique sink API call, and we will also provide the per-app SSG in the future.
\myfig~\ref{fig:exampleBSG} shows an example SSG (simplified for readability) that is automatically generated by \backdroid for the app \path{com.studiosol.palcomp3}.
Compared with traditional slices, our SSG contains the following additional information:

\begin{compactitem}
  \item \textit{Hierarchical taint map.}
    Although not displayed in \myfig~\ref{fig:exampleBSG}, a hierarchical taint map is actually maintained during our inter-procedural backtracking.
    Specifically, our SSG assigns a taint set to each tracked method and organizes all sets hierarchically according to their method signatures.
    We also maintain a global taint set for static fields.
 
  \item \textit{Inter-procedural relationships.}
    Each time our bytecode search uncovers an inter-procedural relationship, we will record it into SSG as a cross-method edge.
    Recall that in the running example of basic search in \mysec\ref{sec:basicSearch}, we connect an edge from the caller (site) to the callee method in the fifth step after the search is done.
    Besides this kind of common cross-method edges, it is also possible for a tracked method to invoke its \textit{contained method}, e.g., \path{MP3LocalServer.<init>()} in \myfig~\ref{fig:exampleBSG}.
    We use both calling and return edges for this special relationship.

  \item \textit{Raw typed bytecode statements.}
    Lastly, to recover a complete representation of sink parameters in the forward analysis, it is necessary to keep raw typed bytecode instructions in the SSG.
    We thus define a node structure called \texttt{SSGUnit} to wrap the original bytecode statements in Soot's \texttt{Unit} format~\cite{SootUnit}.
    In this structure, we record the node ID, the signature of corresponding method, and most importantly, the typed bytecode \texttt{Unit} statement.
\end{compactitem}

\textbf{Backward taint analysis over fields, arrays, and contained methods for the SSG generation.}
With the SSG structure defined, we then perform \textit{backward} taint analysis to generate SSGs.
Compared to the forward taint analysis in Amandroid and FlowDroid, backward taint analysis is more difficult because it reverses the normal program execution and thus has no insights on the earlier execution of tainted variables.
This problem is particularly noticeable for fields, arrays, and contained methods, and we specifically handle them as follows.
First, for an instance field to be tainted, e.g., \texttt{r0.<com.studiosol.util.NanoHTTPD: int myPort>} in \myfig~\ref{fig:exampleBSG}, we add not only the instance field itself to the taint set but also its class object (i.e., \texttt{r0}) so that we can trace the same field no matter the class object gets aliased or across method boundaries.
Moreover, when an instance field needs to be untainted, we first remove \texttt{obj.field} from the taint set and further detect whether there are more fields for the same instance.
If there are no other such fields, we remove \texttt{obj} from the taint set as well.
Arrays and \texttt{Intent} objects are handled in a similar way.
Hence, we skip the details here.

When there are static fields in the taint set, we need to also analyze contained methods (e.g., \path{MP3LocalServer.<init>()} in \myfig~\ref{fig:exampleBSG}).
A normal processing is to jump into all contained methods (even when their parameters are not tainted) and analyze them, because we cannot determine whether a contained method uses a tainted static field or not.
Analyzing all contained methods on the backtracking paths certainly slows down the analysis, and we have proposed a more elegant solution.
Specifically, whenever a new static field is tainted, we launch a bytecode search of this field signature to capture all methods that access this particular static field.
Therefore, we only need to analyze the contained methods that are matched with our search results.

\textbf{Adding off-path static initializers into SSG on demand.}
In \mysec\ref{sec:searchStaticInit}, we introduced how to search over \textit{on-path} static initializers.
Here we continue to explain how to accordingly add \textit{off-path} static initializers, i.e., those not in the backtracking paths.
Specifically, after the main taint process is done, if there are still unresolved static fields in the SSG's taint map, we retrieve the corresponding classes and obtain their \texttt{<clinit>} methods that are only implicitly executed by the Java/Android virtual machine.
We then perform the backward taint analysis of these \texttt{<clinit>} methods, and add only relevant statements into a special track of SSG.
\myfig~\ref{fig:exampleBSG} presents such a track for the \path{MP3LocalServer.<clinit>()} method, which captures the value of an unresolved static field, \path{<com.studiosol.palcomp3.MP3LocalServer:} \texttt{int PORT>}.
During the forward analysis, we will first analyze this special static track and then handle the main track of SSG.

\subsection{Forward Constant and Points-to Propagation over SSG}
\label{sec:forwardBSG}

After producing a complete SSG, our forward analysis iterates through each SSG node, analyzes each statement's semantic, and propagates dataflow facts through the constant~\cite{ConstantPropagation91} and points-to~\cite{JavaPointsTo03} propagation during the graph traversal.
We now explain these forward analysis steps over our new SSG structure.

\textbf{Overall traversal process over SSG.}
As mentioned in \mysec\ref{sec:generateBSG}, an SSG includes two tracks, the special static field track and the normal track.
Our traversal always starts with the static field track so that we can resolve fields referred in the normal track.
For each track, we first retrieve a set of tail nodes (e.g., the starting blocks in \myfig~\ref{fig:exampleBSG}) and initialize analysis from each of them.
To record dataflow facts generated by our analysis, we maintain fact maps for each analysis flow, but we use only one global fact map for all static fields.

Whenever we reach at a new SSG node, we perform graph traversal as follows.
First, we determine whether the node is an initial SSG node with a sink API call (e.g., the ending block in \myfig~\ref{fig:exampleBSG}); and if it is, we correlate and output dataflow facts of all tainted parameters.
For a normal SSG node, we first jump into its invoked methods if any.
After that, we analyze the node itself and move to the next node(s).

\textbf{Analyzing and modeling statement semantics.}
During the traversal of each SSG node, we parse its typed bytecode statement and analyze the semantic.
As shown in \myfig~\ref{fig:exampleBSG}, an SSG contains only three kinds of statements to be handled, namely \texttt{DefinitionStmt}~\cite{SootDefinitionStmt} (and its subclass \texttt{AssignStmt}~\cite{SootAssignStmt}), \texttt{InvokeStmt}~\cite{SootInvokeStmt}, and \texttt{ReturnStmt}~\cite{SootReturnStmt}.
We further analyze the statement expression embedded, which can be one of the six kinds of statement expressions, including \texttt{BinopExpr}, \texttt{CastExpr}, \texttt{InvokeExpr}, \texttt{NewExpr}, \texttt{NewArrayExpr}, and \texttt{PhiExpr}.
We then follow these expression instructions to understand their semantics and calculate dataflow facts.
In particular, we mimic arithmetic operations and model Android/Java APIs to handle two complicated expressions, \texttt{BinopExpr}~\cite{SootBinopExpr} and \texttt{InvokeExpr}~\cite{SootInvokeExpr}.


\textbf{Propagating constant and points-to information.}
To facilitate the dataflow propagation, we maintain a fact map to correlate each variable with its dataflow fact.
Propagating constant facts among different variables is easy --- just retrieve the value from an old variable and assign it to a new variable in the fact map.
To propagate object points-to information, we design an object structure called \texttt{NewObj} to preserve the original points-to information along flow paths.
Each \texttt{NewObj} object contains a pointer to its constructor class, a map of member objects (in any class type) and their reference names.
Then we just need to propagate \texttt{NewObj} objects along flow paths so that all corresponding objects being traced can point to the same \texttt{NewObj} object.
Inner members of \texttt{NewObj} can also be updated by checking classes' \texttt{<init>} methods or any other value-assignment statements.
Besides the class objects' points-to information, we define an \texttt{ArrayObj} object to wrap the points-to information of array expression and its array map between indexes and values.


\section{Evaluation}
\label{sec:BKevaluate}

In this section, we evaluate the efficiency and efficacy of \backdroid in analyzing modern apps. 
In particular, we compare \backdroid with Amandroid~\cite{Amandroid14, Amandroid18}, the state-of-the-art Android static dataflow analysis tool.
Both Amandroid and \backdroid support the dataflow analysis of all kinds of sink-based analysis problems, such as API misuse (e.g.,~\cite{CryptoLint13, BrokenFinger18}) and malware detection (e.g.,~\cite{RiskRanker12, MaMaDroid17}).
In contrast, FlowDroid focuses only on the privacy leak detection that involves both source and sink APIs, and thus it is not compared here. 
Nevertheless, by comparing our \backdroid's result presented in this section with the call graph generation result of FlowDroid in \mysec\ref{sec:callgraph}, we find that \backdroid still performs much faster (2.13min v.s. 9.76min, on average).

\subsection{Experimental Setup}
\label{sec:expsetup}

To fairly evaluate both \backdroid and Amandroid, we select two common and serious sink-based problems, crypto and SSL/TLS misconfigurations, which were also recently tested by Amandroid in \cite{Amandroid18}.
In both cases, the root cause is due to insecure parameters.
For example, the ECB mode is used to create the \path{javax.crypto.Cipher} instance~\cite{Study11, CryptoLint13, CDRep16} and the insecure parameter \path{ALLOW_ALL_HOSTNAME_VERIFIER} is used in \path{setHostnameVerifier()}~\cite{SSLbyMalloDroid12, SSLinNonBrowser12, SMVHunter14}.
Note that these two kinds of sink APIs are frequently invoked in our tested apps --- on average, 21 sink API calls in each app of our dataset, as we will show in \mysec\ref{sec:expFurther}.
In this way, we can stress-test the performance of \backdroid even though we are targeting at just two problems.
In the rest of this subsection, we describe the dataset tested, computing environment used, and tool parameters configured.

\textbf{Dataset.}
We use a set of modern popular apps that satisfy two conditions: they (i) have at least one million installs each, and (ii) were updated in recent years.
Specifically, we first select all such 3,178 apps in our app repository (see Table~\ref{tab:appsizechange} in \mysec\ref{sec:motivate}) as our basic dataset.
However, not all of them contain the specific sink APIs, so we \blue{pre-search} them to filter out the apps with all three target sink APIs, namely \path{Cipher.getInstance()}, \path{SSLSocketFactory.setHostnameVerifier()}, and \path{HttpsURLConnection.setHostnameVerifier()}.
This is to help Amandroid avoid the unnecessary analysis, as it has no bytecode search capability.
\red{Hence, the actual performance gap between Amandroid and \backdroid for analyzing all individual apps could be even larger than what we report in this paper.}
Eventually, we use the searched 144 apps for our experiments.
\blue{The average and median size of these apps are 41.5MB and 36.2MB, respectively, while the largest and the smallest are 104.9MB and 2.9MB, respectively.}

\textbf{Environment.}
For the computing environment, we use a desktop PC with Intel i7-4790 CPU (3.6GHZ, eight cores) and 16GB of physical memory.
Note that a memory configuration with 16GB or less is often used in many previous studies, e.g.,~\cite{AppSealer14, AppContext15, DomainSocket16, HSOMiner17}.
To guarantee sufficient memory for the OS itself, we assign 12GB RAM to the Java VM heap space in running Amandroid. 
To demonstrate that \backdroid is not sensitive to the amount of available memory, we use only 4GB (i.e., \texttt{-Xmx4g}).
The OS is 64-bit Ubuntu 16.04, and we use Java 1.8 and Python 2.7 to run the experiments.

\textbf{Tool configuration.}
Both Amandroid and FlowDroid need to configure a set of parameters to balance their performance and precision.
In contrast, \backdroid does not require specific parameter configuration, since we have programmed its maximum capability in the code. 
In this paper, we use the default Amandroid parameters (see its \texttt{config.ini} file), and use the latest Amandroid 2.0.5 that supports the inter-procedural API misuse analysis\footnote{Amandroid after version 2.0.5 uses only the \textit{intra}-procedural dataflow analysis to analyze API misuse, see details at \url{https://github.com/arguslab/Argus-SAF/issues/55}.}.
We also give Amandroid sufficient running time with a large timeout of 300 minutes for each app.

\subsection{Performance Results}
\label{sec:expPerformance}

Out of the 144 apps analyzed, both Amandroid and \backdroid successfully finished the analysis of 141 apps.
For Amandroid, the three failures are all due to its errors in parsing the manifest, while for \backdroid, two failures are caused by the format transformation from bytecode to IR (in \texttt{com.kbstar.liivbank} and \texttt{com.lguplus.paynow}) and one failure is a Soot bug in processing the \texttt{com.lcacApp} app.
In our evaluation, we thus do not count these failed apps.

\begin{figure}[t!]
\begin{adjustbox}{center}
\includegraphics[width=0.4\textwidth]{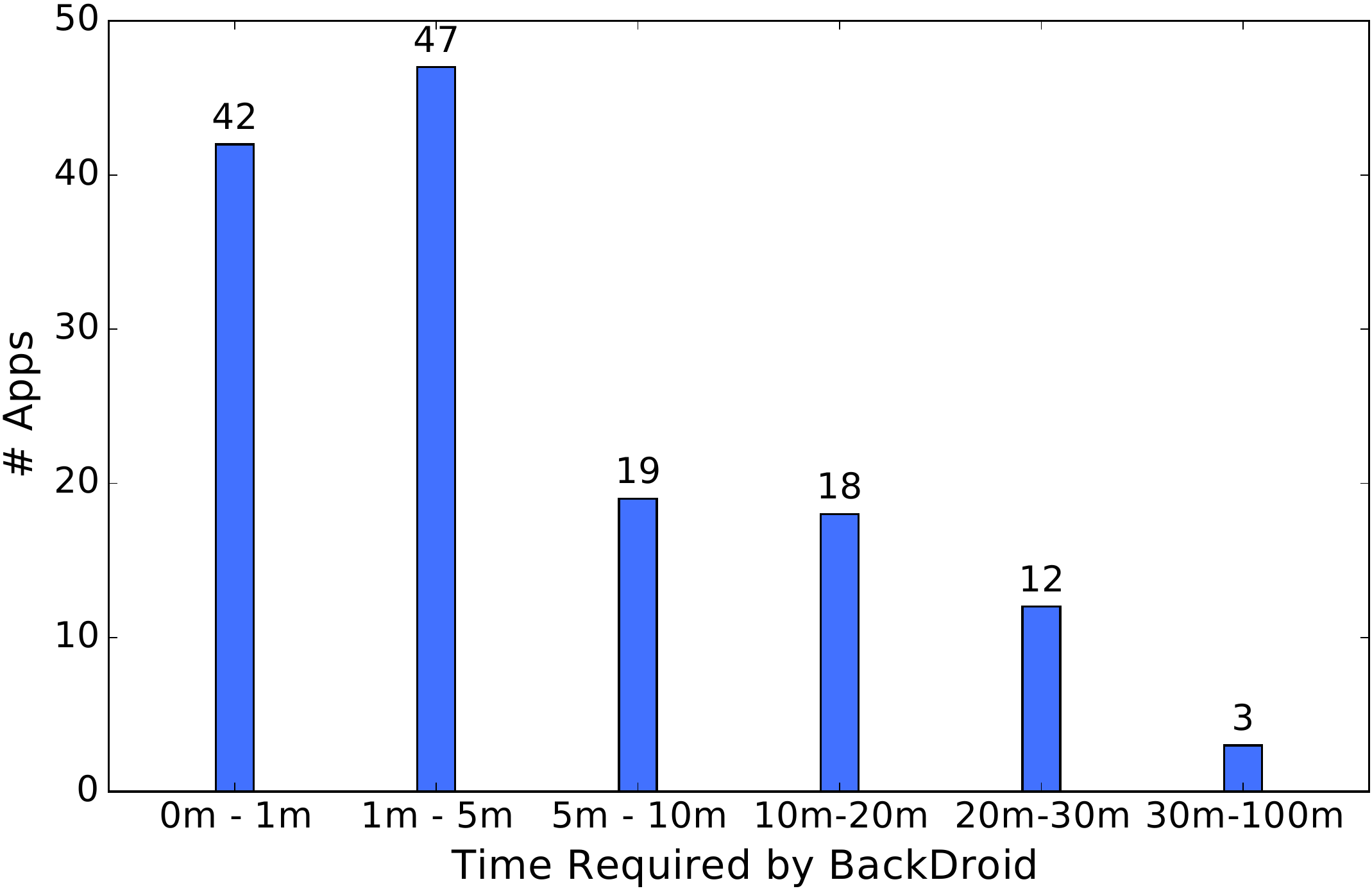}
\end{adjustbox}
\caption{The distribution of analysis time in \backdroid.}
\label{fig:backTimeBar}
\end{figure}

\begin{figure}[t!]
\begin{adjustbox}{center}
\includegraphics[width=0.4\textwidth]{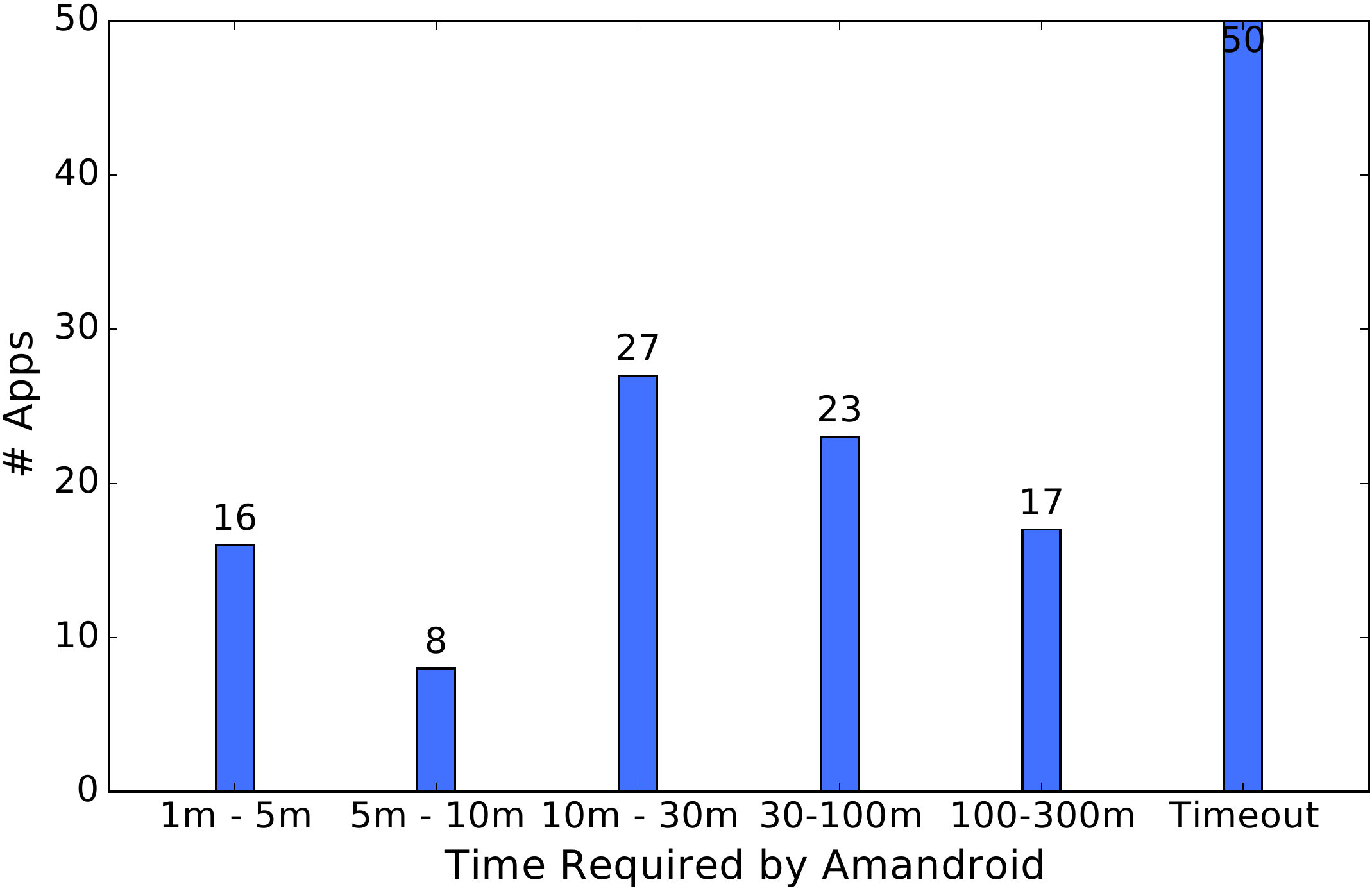}
\end{adjustbox}
\caption{The distribution of analysis time in Amandroid.}
\label{fig:amanTimeBar}
\end{figure}

\myfig~\ref{fig:backTimeBar} and \myfig~\ref{fig:amanTimeBar} show the distribution of analysis time used by \backdroid and Amandroid, respectively.
By correlating these two figures, we make the following three observations on the performance of \backdroid and Amandroid.

First, \textit{\backdroid's has no any timed-out failure, while the timed-out failure rate in Amandroid is as high as 35\%}.
Even though we \blue{have set a considerably large timeout for Amandroid (five hours for every single app)}, there are still as many as 50 apps timed out in Amandroid, as shown in \myfig~\ref{fig:amanTimeBar}.
This 35\% (50/141) timed-out failure rate indicates that Amandroid is out of control in analyzing modern apps.
In contrast, \myfig~\ref{fig:backTimeBar} shows that only three apps in \backdroid exceeded 30 minutes, with 35min, 39min, and 81min, respectively.
This suggests that \backdroid's analysis time is always under control even when we are analyzing a set of large apps.
In \mysec\ref{sec:expFurther}, we will further analyze and show that \backdroid's performance largely depends on the number of sink API calls analyzed, instead of the app/code size that existing tools are mainly affected by (see Amandroid and FlowDroid's own evaluation~\cite{Amandroid14, FlowDroid16Thesis}).


Second, \textit{\backdroid can quickly finish the analysis of most of the apps, with 77\% apps analyzed within 10 minutes and even with one third of apps finished within just one minute}.
After analyzing the cases of long analysis time, we now focus on the apps with short analysis time.
According to \myfig~\ref{fig:backTimeBar}, \backdroid requires just one minute to analyze 30\% (42) apps each, and as high as 77\% (108) apps can be quickly finished within 10 minutes each.
This gives \backdroid a great potential to be deployed by app markets for online vetting.
In contrast, only 17\% (24) apps can be analyzed by Amandroid within the same time period, as shown in \myfig~\ref{fig:amanTimeBar}, and no app could be finished within one minute.

Third, \textit{the overall performance of \backdroid, in terms of the median time, is 37 times faster than that in Amandroid}.
After studying apps that take relatively long and short analysis times, we further analyze the overall performance.
Since there are 50 timed-out failures in Amandroid's result, measuring the average analysis time is not reliable.
Hence, we analyze the median time, and find that the performance gap between the two tools is quite significant: the overall median time of all 141 apps analyzed in \backdroid is 37 times faster than that in Amandroid (2.13min versus 78.15min).

\begin{framed}
\vspace{-2ex}
\small
\noindent
\textbf{Performance Takeaway:}
\backdroid achieves much better efficiency than Amandroid, with the overall performance 37 times faster, and \blue{with no timeout} and much more quick analysis cases.  
\vspace{-2ex}
\end{framed}

\subsection{Detection Results}
\label{sec:expDetection}

After comparing \backdroid's and Amandroid's efficiency, we now analyze and compare their detection accuracy.
We present their detection results from two perspectives.

\textbf{Vulnerabilities detected by Amandroid but not \backdroid.}
We first analyze whether \backdroid could achieve a close detection rate for the app vulnerabilities that are detected by Amandroid.
For the crypto API usage, Amandroid detected that seven apps are still using insecure ECB mode.
We find that \backdroid can accurately detect all of them.
The buggy apps include the popular Adobe Fill \& Sign app (\texttt{com.adobe.fas}) and a bank app called IDBI Bank GO Mobile+ (\texttt{com.snapwork.IDBI}).
Both apps must guarantee a secure encryption in their design.
\red{We have contacted the vendors on these risky security issues.}
We also find that to detect these seven vulnerable apps, Amandroid spent a total of 8.53 hours (73min on average), whereas \backdroid required only 8.78 minutes (1.26min each), around 60 times faster.

Compared to the crypto API misuse, Amandroid detected more SSL misconfigurations in our dataset, with 23 apps discovered with the wrong SSL hostname verification.
However, our diagnosis shows that six of them are false positives.
Among the 17 true positives, \backdroid detected 15 of them and failed on two apps. 
The detailed diagnosis results are:
\begin{compactitem}
  \item \textit{Four false positives from the ArmSeedCheck library}: Amandroid reported four buggy apps with a library called \path{com.skt.arm.ArmSeedCheck}. However, three of them do not trace back to entry functions, and the sink flow of one app (\path{kemco.hitpoint.machine}) originates from an \texttt{Activity} component (\path{jp.kemco.activation.TstoreActivation}) not in manifest.

  \item \textit{Two false positives from the qihoopay library}: Two vulnerable apps with the \path{com.qihoopay.insdk.utils.HttpUtils} library were reported by Amandroid.
    However, their sink flows similarly come from a unregistered and thus deactivated \texttt{Activity} component.

  \item \textit{Two false negatives in \backdroid}: Unfortunately, \backdroid failed on the \path{com.gta.nslm2} and \path{com.wb.goog.mkx} apps.
    The root cause for both cases is that they do not directly invoke the system sink APIs, which makes our initial bytecode search step fail to locate their sink API calls.
    For example, the \path{com.gta.nslm2} app's \path{com.youzu.android.framework.http.client.DefaultSSLSocketFactory} class extends the system API class \path{org.apache.http.conn.ssl.SSLSocketFactory} and invokes the \path{setHostnameVerifier()} API method only via its own method signature. 
    We will address this issue by checking the class hierarchy also in the initial search.
\end{compactitem}

\textbf{Vulnerabilities detected by \backdroid but not Amandroid.}
We further find that for some apps, \backdroid can achieve better detection than Amandroid.
In particular, \backdroid discovered 54 additional apps with potentially insecure ECB and SSL issues that were not detected by Amandroid.
By analyzing them, we identify four important factors \red{(the last two, with 18 apps, are fully due to Amandroid's inaccuracy)}:
\begin{compactitem}
  \item \textit{Timed-out failures}: 28 of the 54 failed apps were due to the timeouts, where Amandroid did not finish their analysis even after running 300 minutes each.

  \item \textit{Skipped libraries}: Recall that Amandroid by default skips the analysis of some popular libraries that are specified in its \texttt{liblist.txt} configuration file.
    For this reason, Amandroid failed to detect the ECB/SSL issues in eight apps, which use the skipped Java packages from Amazon, Tencent, and Facebook, such as \path{com.amazon.identity.frc.helper.EncryptionHelper} and \path{com.tencent.smtt.utils.LogFileUtils}.

  \item \textit{Unrobust handling of certain implicit flows}: 
    We surprisingly find that Amandroid is not as robust as \backdroid to handle certain asynchronous flows and callbacks.
    For example, Amandroid failed to connect the flow from \path{AsyncTask.execute()} to \path{doInBackground()} and the callback from \path{setOnClickListener()} to \path{onClick()} in some apps.
    Such unrobust handling appeared in eight of the 54 apps.

  \item \textit{Occasional errors in its whole-app analysis}: By inspecting Amandroid's debug logs, we observed some occasional errors (e.g., ``Could not find procedure'' and ``key not found'') during the analysis of Amandroid, which cause 10 apps fail to be detected.
    By nature, it is easier for Amandroid's whole-app analysis to encounter errors, as compared with \backdroid's targeted analysis.
\end{compactitem}


\begin{framed}
\vspace{-2ex}
\small
\noindent
\textbf{Accuracy Takeaway:}
\backdroid achieves close detection effectiveness for the apps that can be detected by Amandroid, and obtains better detection results for the apps with long analysis time, skipped libraries, and certain asynchronous flows/callbacks.  
\vspace{-1ex}
\end{framed}


\subsection{Further Measurement}
\label{sec:expFurther}

After presenting performance and detection results, we further quantify the relationship between the number of sink API calls and \backdroid's analysis time.
This is to answer the concern whether \backdroid could keep its performance when analyzing a large number of sink API calls.
Note that we have tried to simulate such situation by targeting at the sink APIs that are frequently invoked, as explained in \mysec\ref{sec:expsetup}.
On average, each app in our dataset contains 20.93 sink API calls analyzed, the number of which should be enough to simultaneously cover many other uncommon sink APIs like \texttt{sendTextMessage()}~\cite{DroidRanger12}, \texttt{ServerSocket()}~\cite{OpenPort19}, and \texttt{LocalServerSocket()}~\cite{DomainSocket16}.

\begin{figure}[t!]
\begin{adjustbox}{center}
\includegraphics[width=0.4\textwidth]{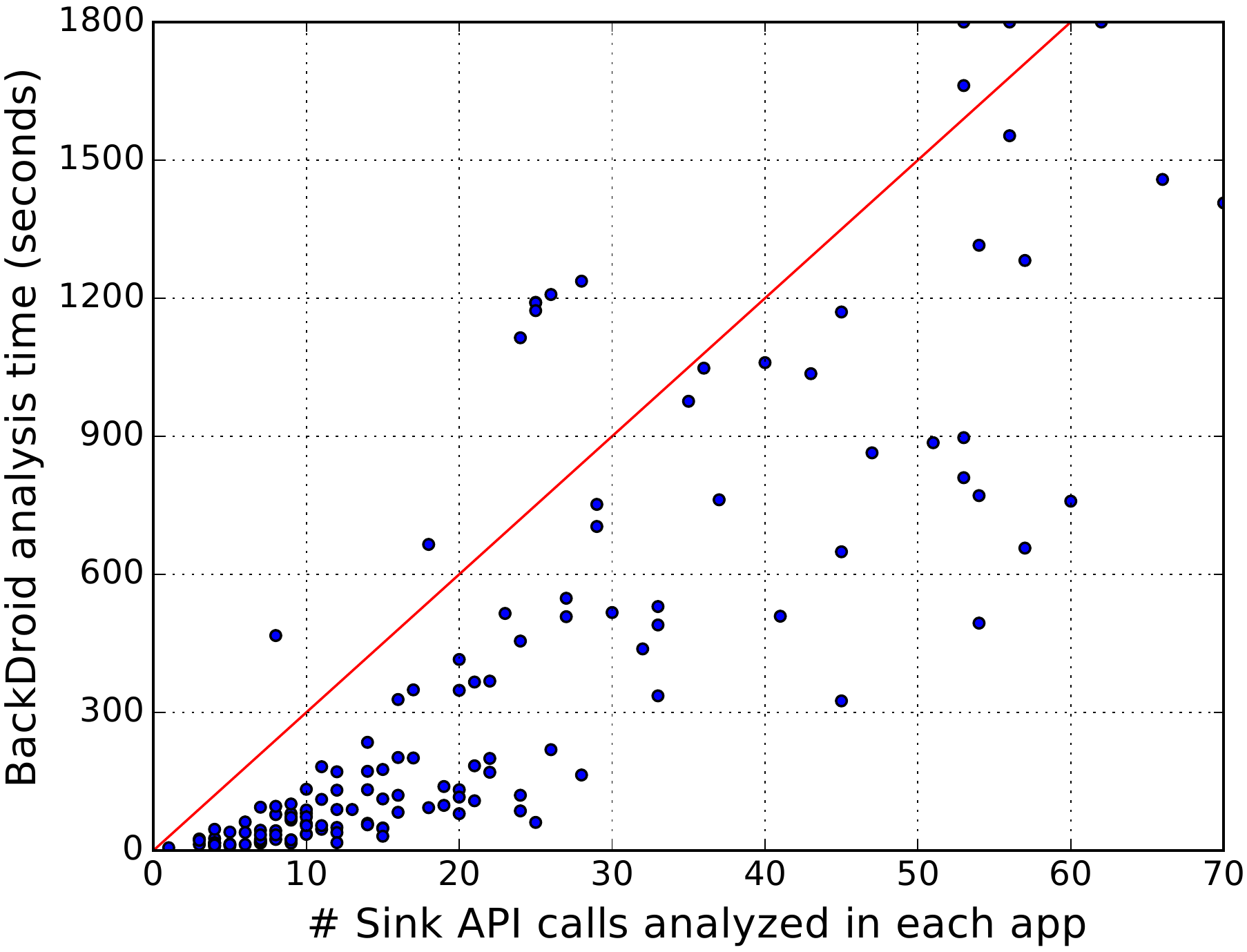}
\end{adjustbox}
\caption{The relationship between the number of sink API calls in each app and \backdroid's analysis time.}
\vspace{-4ex}
\label{fig:timeWithSinkNum}
\end{figure}

In \myfig~\ref{fig:timeWithSinkNum}, we further plot the relationship between the number of sink API calls analyzed in each app and \backdroid's analysis time.
We can see that the majority of them are at a speed faster than 30 seconds per sink call.
For example, we can draw a straight line from the coordinate (0, 0) to the coordinate (60, 1800), and only around ten dots are above this line.
Under this linear trend, \backdroid is expected to finish the analysis of 100 sink calls with around 50 minutes generally.
Indeed, all apps, except one, were analyzed within 40 minutes.
The only outlier, Huawei Health, costed 81min for its 121 sink API calls, which is still much faster than the 300-minute timeout we assigned to Amandroid.
This suggests that \backdroid's analysis time could be always under control.


To further improve \backdroid in analyzing a large number of sink API calls, we will evolve the current per-sink SSG to per-app SSG, as mentioned in \mysec\ref{sec:generateBSG}.
In this way, we can guarantee that no matter how many sinks there are, \backdroid only requires to generate a \textit{partial}-app graph once, which would be still more efficient than existing whole-app graphs.

\section{Discussion}
\label{sec:BKdiscuss}



So far, we have elaborated our approach in the context of Android bytecode.
There are some common technical issues in typical Android app analysis works, namely Java reflection, native code, dynamically loaded code, and packed code.
Although addressing these issues is not our focus in this paper, we discuss our plan to mitigate them in the future work.

\textbf{Java reflection.}
Java reflection is a mechanism that can dynamically and secretly invoke an Java API by setting different method parameters in reflection API calls.
To mitigate such effect for static analysis, an immediate solution is to leverage DroidRA~\cite{DroidRA16} to transform the original app to a version without reflection calls.
In the future, we will first resolve reflection parameters using our on-the-fly backtracking and then directly build caller edges to cache them.

\textbf{Native code.}
To extend \backdroid's design principle also to native code~\cite{PushPopRandom15, GoingNative16} and JNI (Java Native Interface)~\cite{JGRE17}, a potential way is to replace \texttt{dexdump} with \texttt{objdump}.
Searching over native code text might be different from bytecode search, and we will explore it.
Nevertheless, given the small size of native code in Android apps and their limited entry points, it is possible to launch full-scale forward analysis, as demonstrated in the recent SInspector~\cite{DomainSocket16} and JN-SAF~\cite{JNSAF18}.

\textbf{Dynamically loaded and packed code.}
Although static analysis theoretically cannot handle dynamically loaded~\cite{CodeLoad14} or packed~\cite{DexHunter15} code, in practice, we can first leverage dynamic analysis~\cite{StaDynA15, TIRO18} to extract those hidden code before running \backdroid.
Note that WebView code~\cite{WebViewAtk11, FileCross14, IndirectFileLeak15, CrossAppWebViewInfection17} could be handled in a purely static manner~\cite{BridgeScope17, DCVHunter19, OAuthLint19}.

\red{Besides these technical improvements, we can also extend the current scope of \backdroid (i.e., targeted vetting of sink-based API usages) and generalize its on-the-fly bytecode search technique to other security problems involving both sources and sinks, notably the privacy leak detection problem.
Specifically, we can first employ on-the-fly backward analysis to determine the reachability of a source API call by tracing from sources to entry points, and then launch on-demand forward dataflow analysis starting only for those reachable source calls to determine whether there is a leak from source to sink.
Since FlowDroid does not support sink-based API misuse detection and \backdroid in this paper does not aim to analyze privacy leaks, we thus do not compare them (except a rough performance comparison shown earlier in \myfig\ref{fig:flowTimeBar} and \ref{fig:backTimeBar}).}

\section{Related Work}
\label{sec:related}


\red{With the background of Android static analysis introduced in \mysec\ref{sec:background}, we now discuss more related works on the topics of search-based program analysis, more scalable, and more accurate Android static analysis in this section.}

To the best of our knowledge, we are the first to introduce the search-based analysis to Android static program analysis.
We also study whether such a concept has been proposed in other domains.
The closest one is a work called ``Search-Based Program Analysis''~\cite{SearchBasedProgramAnalysis11}, which was indeed a keynote speech on the combination of search-based test case generation~\cite{SearchBasedTest10, SymbolicSearchTest11} and program analysis.
Besides this software engineering concept, security researchers leveraged conditional searches as database queries to speed up the bug detection~\cite{Genius16, CoddPropertyGraph14} and network security discovery~\cite{Censys15} at a large scale.
\red{There are two major differences between these works and our on-the-fly bytecode search.
First, our search is not a conventional search like a graph/database/text search, but rather a bytecode semantic-aware search.
Second, we are the only one that deeply embeds a code-level search into program analysis.}

Some recent works~\cite{HSOMiner17, BrokenFinger18, FlowCog18} have also realized the scalability difficulty in existing Android static analysis.
For example, because of the worry that it is slow to launch complicated code analysis in a set of apps with the average size of 8.43MB, HSOMiner~\cite{HSOMiner17} proposed to combine lightweight program analysis with machine learning for the large-scale detection of hidden sensitive operations.
Similarly, in a recent work by Bianchi et. al.~\cite{BrokenFinger18}, it stated the challenge in analyzing recent real-world Android apps due to their large amount of code. 
As a result, they performed \textit{intra}-procedural type-inference analysis to construct their whole-app call graph.
\red{More recently, to speed up the cryptographic vulnerability detection for massive-sized Java projects, CryptoGuard~\cite{CryptoGuard19}, proposed the crypto-specific slicing that uses a set of domain knowledge on top of the intra-procedural dataflow analysis of Soot to balance the precision and runtime tradeoff.}

\red{Besides the performance, accuracy is a top concern of many Android static analysis tools.
Besides the widely-used Amandroid and FlowDroid, DroidSafe~\cite{DroidSafe15} and HornDroid~\cite{HornDroid16} are two representatives towards more accurate Android static analysis.
Specifically, DroidSafe provides a comprehensive model of the Android runtime via a technique called \textit{accurate analysis stubs} and combines it with carefully designed inter-procedural and ICC analysis.
We also plan to incorporate such analysis stubs into the forward object taint analysis part of \backdroid (see \mysec\ref{sec:advancedSearch}).
On the other hand, HornDroid is essentially a symbolic execution tool with SMT constraint solving. 
Despite these great efforts, a recent work, $\mu$SE~\cite{uSE18}, found that technical inaccuracy still exists.
Specifically, it used systematic mutation to discover flaws, e.g., missing callbacks and incorrect modeling of asynchronous methods, in all of those four tools.}
In our evaluation of Amandroid in \mysec\ref{sec:expDetection}, we also identified certain callbacks and asynchronous flows missed by Amandroid.
Besides $\mu$SE, two surveys~\cite{DroidTools16, DroidTools16TSE} also tried to systematically assess the performance and accuracy of existing Android static analysis tools.

\section{Conclusion}
\label{sec:BKconclude}

In this paper, \red{we proposed a new paradigm of targeted inter-procedural analysis by combining traditional program analysis with our on-the-fly bytecode search.
We implemented this technique into a tool called \backdroid for targeted and efficient security vetting of modern Android apps.
We overcame unique challenges to search over Java polymorphism, asynchronous flows, callbacks, static initializers, and Android inter-component communication, and also adjusted the traditional backward slicing and forward analysis over a structure called self-contained slicing graph (SSG) for the complete dataflow tracking.}
Our experimental results showed that \backdroid is 37 times faster than Amandroid and has no timeout (v.s. 35\% in Amandroid), while it maintains close or even better detection effectiveness.
In the future, we will enhance \backdroid to search over Java reflection and native code, and also extend it to other non-sink-based problems like privacy leak detection. 




%
\bibliographystyle{IEEEtranS}   
\bibliography{main}

\end{document}